\documentclass[11pts]{LMA}

\newcommand{\rd}{R_{D}}
\newcommand{\rds}{R_{D^{*}}}
\newcommand{\rdsb}{R_{D^{(*)}}}
\newcommand{\rksb}{R_{K^{(*)}}}

\definecolor{mo}{HTML}{c47207}
\newcommand{\cfg}{\color{ForestGreen}} 
\newcommand{\cv}{\color{violet}}
\newcommand{\cs}{\color{mo}}

\begin{document}
\title{Common explanation to the $\rksb$, $\rdsb$ and $\epsilon^\prime/\epsilon$ anomalies in a 3HDM+$\nu_R$ and connections to neutrino physics
	\vspace{.3em}}

\author[]{Carlo Marzo
\thanks{carlo.marzo@kbfi.ee (corresponding author)}
}

\author[]{Luca Marzola
\thanks{luca.marzola@cern.ch}
}

\author[]{Martti Raidal
\thanks{martti.raidal@cern.ch}
}

\affil[]{\KBFI}

\date{\today}
\maketitle

\begin{abstract}
	Scalar theories can account for the current $\rdsb$ measurements through a  vector operator $\bar{c}_L \gamma_{\mu} b_L\,\bar{\tau}_L \gamma^{\mu}\nu_L$ induced at the loop level. Once the vector contribution is considered on top of a subdominant tree-level scalar component, the predicted value of $\rdsb$ falls within the $1\sigma$ region indicated by the experiments. We explicitly demonstrate this claim in the framework of a three Higgs doublet model extended with GeV scale right-handed neutrinos, by matching the anomalous signal for perturbative values of the involved couplings and respecting the bounds from complementary flavour physics measurements. Remarkably, we furthermore show that the proposed framework can be employed to simultaneously explain also the present $\rksb$ measurement, as well as the deviation in $\epsilon'/\epsilon$ currently being debated in the literature. These results are obtained by considering the contribution of relatively light right-handed neutrinos which are fundamental in mediating the processes behind the anomalous signals. In this way our findings reveal a new possible connection that links the flavour anomalies to the phenomenology of extended Higgs sector and neutrino physics.
\end{abstract}
\vspace{1 cm}
\section{Introduction} 
\label{sec:Introduction}

In recent years, $B$ factories and the LHC$b$ experiment have reported several anomalies~\cite{Aaij:2015yra,Aaij:2017uff,Lees:2012xj,Lees:2013uzd,Huschle:2015rga,Sato:2016svk,Hirose:2016wfn,Aaij:2014ora, LHCB,ATLAS:2017dlm,CMS:2017ivg} that find no satisfying explanation in the flavour structure and interactions supported by the standard model (SM)~\cite{Capdevila:2017bsm,Altmannshofer:2017yso,DAmico:2017mtc,Hiller:2017bzc,Geng:2017svp,Ciuchini:2017mik,Becirevic:2017jtw,1592392,DiChiara:2017cjq,Li:2016vvp,Calibbi:2017qbu, Megias:2016bde,Blanke:2018sro,Asadi:2018wea,Greljo:2018ogz,Abdullah:2018ets,Iguro:2018qzf,Fraser:2018aqj}. The interpretation of these signals in term of new physics faces, as well, several difficulties, in particular the simultaneous explanation of the $\rksb$ and $\rdsb$ measurements proves undoubtedly challenging for the implications on further precision observables that rule out the underlying proposals~\cite{Crivellin:2012ye,Iguro:2017ysu,Lee:2017kbi,Alonso:2016oyd,Akeroyd:2017mhr,Martinez:2018ynq,Jung:2018lfu,Kumar:2018kmr}.

These difficulties are grounded in the different origin of these anomalies, which seemingly require new contributions to appear at several scales and put into discussion properties, such as the universality of gauge couplings, thoroughly verified in collider experiments. For instance, since $\rdsb$ quantifies the ratio of processes that receive tree-level contributions within the SM, the measured value of this parameter naturally calls for new degrees of freedom below the electroweak scale. On the contrary, as $\rksb$ is purely determined by the loop structure of the SM, we naively expect that deviations in this observable originate at a much larger scale barring the fine-tuning of the involved couplings. Furthermore, the anomalous signals require new sources of lepton flavour universality violation which, in concrete frameworks, are generally tightly constrained by collider and complementary flavour experiments. 

If we analyse the SM case, we see that the only source of flavour non-universality is the Higgs sector, where the Yukawa couplings of fermions necessarily reflect the measured mass hierarchy. Consequently, extending the Higgs sector could be a natural way to implement the additional sources of flavour non-universality at the basis of the measured signals. 

In the present paper we pursue this possibility, showing that the $\rksb$ and $\rdsb$ measurements can simultaneously be addressed with an extended Higgs sector. The result is based on the preliminary investigation in Ref.~\cite{Fraser:2018aqj}, where we demonstrated how scalar extensions of the SM can generate a sizeable vector operator $\bar{c}_L \gamma_{\mu} b_L\,\bar{\tau}_L \gamma^{\mu}\nu_L$ at the loop level. Explaining the mentioned flavour anomalies in respect of the complementary experimental bounds requires at least two new scalar particles, which must belong to different gauge multiplets, as well as the inclusion of right-handed neutrinos (RHNs). A natural ultraviolet completion of the proposed framework is then provided by a three Higgs doublet model (3HDM), augmented with RHNs, which we analyse in detail in the following. 

As we will show, the inclusion of the tree-level scalar operators contributions allows the model to reproduce the anomalous $\rdsb$ signal for values of the involved couplings of order unity. We consider all the emerging effective chiral operators, including the ones mediated by relatively light right-handed neutrinos usually omitted in these studies. We explicitly propose the Yukawa coupling texture of the additional Higgs doublets needed to explanation the anomaly. Remarkably, we find that the same texture allows also for the explanation of the $\rksb$ anomaly owing to the presence of Majorana RHNs, used here exclusively to mediate the processes behind these signals\footnote{Alternative explanations of the anomalous signals which rely on the presence of RHNs in the involved final states can be found in Refs.~\cite{Greljo:2018ogz,Asadi:2018wea,Babu:2018vrl,Azatov:2018kzb,Robinson:2018gza}}. In this light, flavour anomalies could therefore constitute a first indirect collider signature supporting the existence of RHNs. The proposed framework then connects the issue of flavour measurements to other open questions within neutrino physics and cosmology, such as the neutrino mass generation mechanism and the dynamics of baryogenesis via leptogenesis, extending the phenomenological reach of flavour experiments. We highlight that RHNs in the considered mass range can be discovered by the planned SHIP experiment at CERN~\cite{Alekhin:2015byh}, as well as in collider experiments through the implied lepton number violating effects~\cite{Antusch:2017ebe}.

A brief part of the present paper is dedicated to the assessment of a further potential flavour anomaly, connected to the amount of direct CP violation measured in $K \to \pi \pi$ decays and quantified in the ratio $\epsilon^\prime/\epsilon$. In this case, without advocating for the necessity of a new physics contribution, we find that a minimal modification of the proposed Yukawa texture would allow to explain the signal, should it survive further scrutiny by the dedicated community.  

The paper is organized as follows. Section~\ref{sec:WC} is dedicated to the $\rdsb$ anomaly: after a brief review of its experimental status, we show how the considered framework explains the anomalous signal. The $\rksb$ anomaly is studied in a similar fashion in Sec.~\ref{sec:WCRK}. In Sec.~\ref{sec:Discussion} we then present the result of our investigation, analyse the compatibility of the two anomalies and detail the potential impact of new determinations of $\epsilon^\prime/\epsilon$.  Finally, our conclusions are presented in Sec.~\ref{sec:Conclusions}.

\section{The $\rdsb$ anomaly} 
\label{sec:WC}

We start by briefly reviewing the present experimental status of the $\rdsb$ anomaly, setting out the formalism at the basis of the analysis. 

\subsection{Experimental status and effective Lagrangian} 
\label{sub:Experimental Status}

The results of the LHC$b$ experiment~\cite{Aaij:2015yra,Aaij:2017uff} have underlined the presence of lepton flavour violating dynamics in the charged decays of the $B$-meson. In more detail, anomalies have been measured in the ratio of branching fractions
\begin{equation} 
R_{D^{(*)}} = \frac{\mathcal{B}(\bar{B} \to D^{(*)} \tau \bar{\nu})}{
\mathcal{B}(\bar{B} \to D^{(*)} \ell \bar{\nu})} \,,
\end{equation}
for $\ell=e,\mu$. Being defined as ratios, $\rd$ and $\rds$ are observables of particular importance that test lepton universality in the transition $b \to c \tau \bar{\nu}_\tau$ net of the involved hadronic uncertainties. 

If considered along the initial measurements by the BaBar~\cite{Lees:2012xj,Lees:2013uzd} and Belle~\cite{Huschle:2015rga,Sato:2016svk,Hirose:2016wfn} collaborations, the LHC$b$ observations set~\cite{HFAG}
\begin{eqnarray}
R_{D}^{\mathrm{exp}} &=& 
0.407 \pm 0.039 \pm 0.024 \,,
\nonumber \\
R_{D^*}^{\mathrm{exp}} &=& 
0.306 \pm 0.013 \pm 0.007 \,,
\end{eqnarray}
highlighting a strong affinity for third generation leptons. The implied departure from universality falls well beyond the SM predictions
\begin{eqnarray}
R_{D}^{\mathrm{SM}} &=& 
0.300 \pm 0.008 \,,
\nonumber \\
R_{D^*}^{\mathrm{SM}} &=& 
0.252 \pm 0.003 \,,
\end{eqnarray}
with a combined significance of about $4\sigma$~\cite{HFAG}.  

The possible presence of new physics in $B$ anomalies is investigated by detailing the low-energy regime of a potential model, through the construction of an effective theory characterized by dimension six operators that preserve color and electric charge. Then, at the $b$-quark mass scale, the theory is described by the effective Lagrangian~\cite{Goldberger:1999yh,Cirigliano:2012ab,Dutta:2017wpq,Bardhan:2016uhr}
\begin{eqnarray} \label{effL}
\mathcal{L}_{eff}^{b \rightarrow c \ell \bar{\nu}} &=& -\frac{2 G_F V_{cb}}{\sqrt{2}} \, \left( C^\ell_{VL}\,\mathcal{O}^\ell_{VL} + C^\ell_{AL}\,\mathcal{O}^\ell_{AL} + C^\ell_{SL}\,\mathcal{O}^\ell_{SL} + C^\ell_{PL}\,\mathcal{O}^\ell_{PL} \right) \,, 
\end{eqnarray}
where $\ell = e, \, \mu, \,\tau$. The four effective operators that appear above are given by
\begin{eqnarray} \label{basis}
\mathcal{O}^\ell_{VL} &=& \left[\bar{c} \gamma^{\mu} b\right]\left[\bar{\ell}\gamma_{\mu} P_L \nu_{\ell}\right]\,, \\
\mathcal{O}^\ell_{AL} &=& \left[\bar{c} \gamma^{\mu} \gamma_5 b\right]\left[\bar{\ell}\gamma_{\mu} P_L \nu_{\ell}\right] \,, \\
\mathcal{O}^\ell_{SL} &=& \left[\bar{c} b\right]\left[\bar{\ell} P_L \nu_{\ell}\right]\,, \\
\mathcal{O}^\ell_{PL} &=& \left[\bar{c} \gamma_5 b\right]\left[\bar{\ell} P_L \nu_{\ell}\right] \,. 
\end{eqnarray}
In this formalism, the dynamics of the $b \rightarrow c \ell \bar{\nu}$ process is fully encoded in a corresponding set of Wilson coefficients $\boldsymbol{C}^\ell = C^\ell_{VL}, C^\ell_{AL}, C^\ell_{SL}, C^\ell_{PL}$, which parametrize the relevant observables through the involved branching fractions
\begin{eqnarray}
\rdsb = \frac{\mathcal{B}_{\tau}^{D^{(*)}}(m_{\tau},\mathbf{C}^{\tau})}{\mathcal{B}_{\mu/e}^{D^{(*)}}(m_{\mu/e},\mathbf{C}^{\mu/e})}\,.
\end{eqnarray} 

The contributions of heavy states to the effective Lagrangian in eq.~\eqref{effL} is computed by matching the full theory to the effective one at the scale where the heavy degrees of freedom are integrated out. The final expression of the Lagrangian is then obtained upon the computation of the RGE evolution down to the scale at which the process is probed. The matching has been performed diagrammatically and we checked that the QCD and QED corrections induced by the RGE evolution have a negligible impact on the couplings in eq.~\eqref{yuktex}. 

\begin{figure}[t]
	\centering
	\includegraphics[width=.42\linewidth]{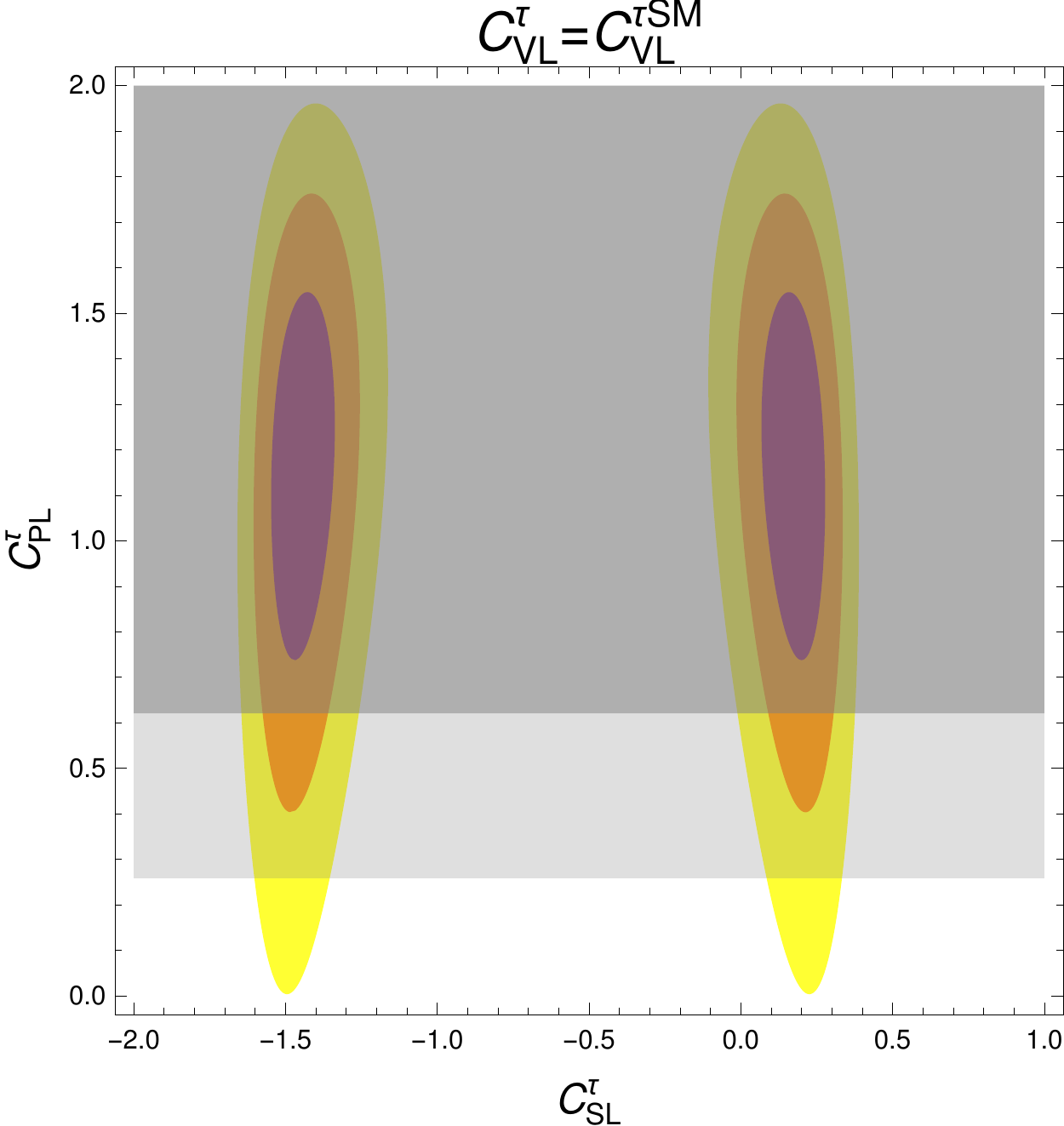}
	\hspace{.5cm}
	\includegraphics[width=.42\linewidth]{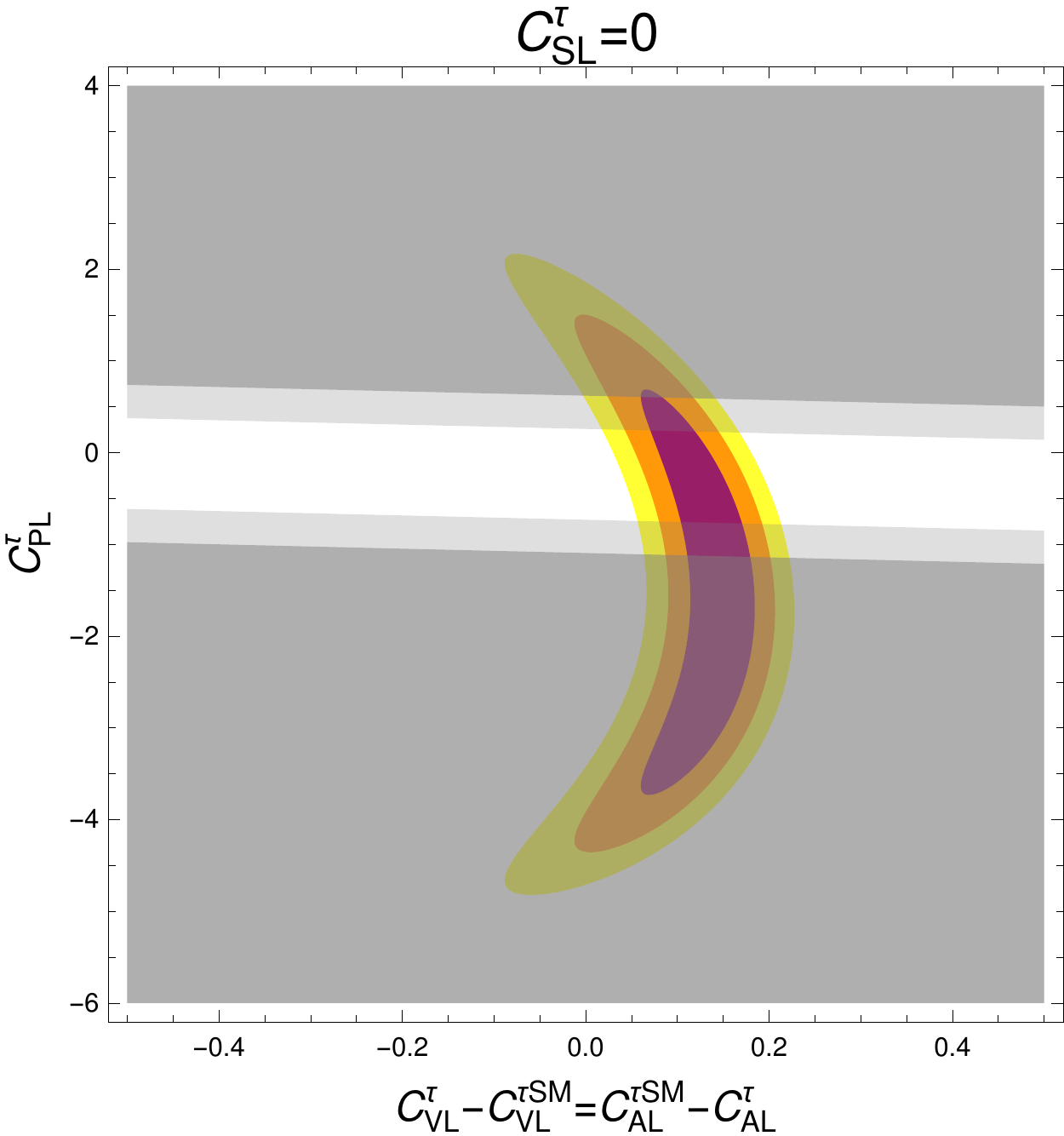}
	\caption{Model-independent fit of the anomalous signal as a function of the indicated Wilson coefficients from eq.~\eqref{effL}. The 68\%, 95\% and 99\% confidence intervals selected by the joint fit of $\rd$ and $\rds$ are marked by the red, orange and yellow areas, respectively. The shaded light (dark) gray areas show instead the current bound from the measured $B_c$ lifetime assuming a 10\% (30\%) maximal allowed deviation.}
	\label{fig1}
\end{figure}

A general analysis of Wilson coefficients provides a first portray of the framework we seek. For instance, the simplest scalar extensions of the SM yield tree-level contributions to the scalar and pseudoscalar Wilson coefficients $C^\tau_{SL}$ and $ C^\tau_{PL}$ which, in principle, allow to explain the anomalous $B$-physics signal. As shown in the left panel of Fig.~\ref{fig1}, new contributions to these quantities let the theoretical predictions enter the 68\% confidence interval associated to the signal, denoted by the red areas in the plot. Such a solution, however, is invalidated by measurements of the $B_c$ lifetime, which severely constrain the pseudoscalar Wilson coefficient due to the mass hierarchy of the SM:
\begin{align}
\mathcal{B}_{\tau\nu}&=\,\frac{m_{B_c}m_\tau^2 f_{B_c}^2G_F^2 |V_{cb}|^2}{8\pi \, \Gamma_{B_c^-}}\left(1-\frac{m_\tau^2}{m_{B_c}^2}\right)^2
\left|\frac{m_{B_c}^2}{m_\tau(m_b+m_c)}C^\tau_{PL}- C^\tau_{AL}\right|^2 \,.
\label{Bctaunu} 
\end{align}   
The impact of this constraint is represented in both the panels of Fig.~\ref{fig1} by the areas shaded in light and dark gray, which indicate the values of the Wilson coefficients that result in deviations larger than $10\%$ or $30\%$ from the measured $B_c$ lifetime, respectively. As we can see, solutions characterized by large values of $C^\tau_{PL}$ are obviously disfavored.

As originally proposed in Ref.~\cite{Fraser:2018aqj}, a possible way for scalar extensions to cope with the $B_c$ lifetime constraint is to rely on scalar loop contributions to the vector and pseudovector Wilson coefficients. The case is shown in the right panel of Fig.~\ref{fig1}, which makes clear that modest values of $C^\tau_{VL}$ and $C^\tau_{AL}$ allow to explain the observed anomaly even when the strongest constraint on the $B_c$ lifetime is considered. A potential issue with this solution is that contributions to $C^\tau_{VL}$ and $C^\tau_{AL}$ are generated in scalar theories only at the loop-level. The required magnitude, of about $\mathcal{O}{(10^{-1})}$, may then impose couplings on the verge of non-perturbativity~\cite{Fraser:2018aqj}. 

We address here this problem by explaining the observed violation of lepton universality through vector and pseudovector contributions on top of a subdominant pseudoscalar component, used to relax the values of the coupling involved in the loop diagrams that generate $C^\tau_{VL}$ and $C^\tau_{AL}$. The interplay between the involved Wilson coefficients is shown in the right panel of Fig.~\ref{fig1}, where crescent values of $C^\tau_{PL}$ allow an excellent fit of the anomaly for lower values of the vector and pseudovector contributions.

\subsection{The 3HDM contribution to $\rdsb$} 
\label{sub:Explaining $\rdsb$ in the 3HDM}

In order to respect the tight bounds on $B\rightarrow X_s \gamma$, as well as further constraints imposed by penguin diagrams, we follow the setup of Ref.~\cite{Fraser:2018aqj} and consider three different scalar $SU(2)$ doublets 
\begin{eqnarray}
H_0 = \left(\begin{array}{c}
H_0^+ \\
H_0^0
\end{array} \right)\,,\,\,\,
H_1 = \left(\begin{array}{c}
H_1^+ \\
H_1^0
\end{array} \right)\,,\,\,\,
H_2 = \left(\begin{array}{c}
H_2^+ \\
H_2^0
\end{array} \right)\,,
\label{doublets}
\end{eqnarray}
with $H_0$ being the SM Higgs doublet. We arrange the scalar potential so that the new doublets do not develop vacuum expectation values, and restrict their masses to the $\sim 300-350$ GeV range to remain within the reach of current collider searches. As mentioned before, we also consider three RH neutrinos $\nu_R$ with corresponding Majorana mass terms. In regard of this, we require that $m_b - m_c - m_{\tau} \lesssim m_{\nu_R^{\tau}}$ to prevent the related $b$ decay channel, maintaining however $m_{\nu_R^i} << m_t$ to retain sizeable loop contributions. From the neutrino physics point of view, these additional states are therefore sufficiently heavy to decouple from the low-energy dynamics of active SM neutrinos.

The interactions of the new scalar states are detailed in the following Lagrangian 
\begin{equation}
	- \mathcal{L} \supset 
	\bar{Q}_L \tilde{H}_1 \mathcal{Y}_1^u u_R 
	+ 
	\bar{Q}_L \tilde{H}_2 \mathcal{Y}_2^u u_R 
	+ 
	\bar{L}_L H_1 \mathcal{Y}_1^{\ell} \ell_R 
	+ 
	\bar{L}_L H_2 \mathcal{Y}_2^{\ell} \ell_R 
	+ 
	\bar{L}_L \tilde{H}_1 \mathcal{Y}_1^{\nu} \nu_R 
	+ 
	\bar{L}_L \tilde{H}_2 \mathcal{Y}_2^{\nu} \nu_R 
	+ 
    \bar{Q}_L H_1 \mathcal{Y}_1^d d_R + h.c. \, ,
	\label{SU2}
\end{equation}
where, for sake of minimality, we set $\mathcal{Y}_2^d = 0$.

\begin{figure}[h]
	\centering
	\includegraphics[scale=0.95]{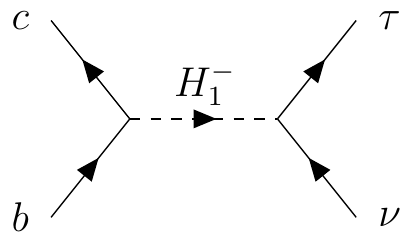}\hspace{.5cm}
	\includegraphics[scale=0.7]{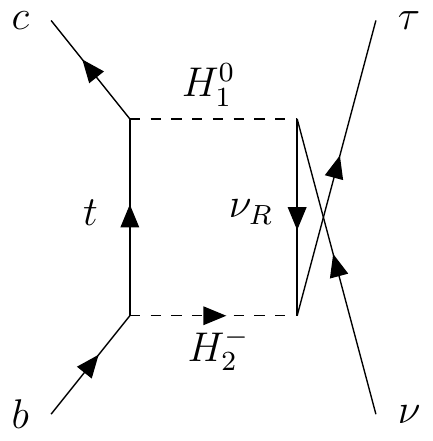}\hspace{.5cm}
	\includegraphics[scale=0.7]{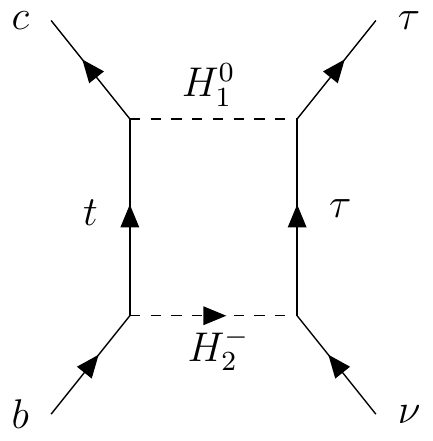}
	\caption{Additional diagrams for the process $b \rightarrow c \tau \bar{\nu}$ supported by the considered 3HDM. The first diagram sources the tree-level expression for $C^\tau_{SL}$ and $C^\tau_{PL}$ given in eqs.~\eqref{cc3} and~\eqref{cc32}. The last two diagrams, instead, yield the contributions to $C_{VL}^\tau$ and $C_{AL}^\tau$ reported in eqs.~\eqref{cc1} and~\eqref{cc2}. We denoted with the symbol $\nu$ an active SM neutrino. }
	\label{fig:S4}
\end{figure}

The Yukawa texture that we investigate is of the form
\begin{eqnarray}
\label{yuktex}
\mathcal{Y}_1^u =\left( \begin{matrix}
	0	& 0 & 0 \\ 
	0	& 0 & {\cfg f_{\bar{c}_L t_R}} \\ 
	0	& {\cv f_{\bar{b}_L c_R}} & 0 
\end{matrix}\right)\,,
\quad
\mathcal{Y}_1^d =\left( \begin{matrix}
0	& 0 & 0 \\ 
0	& 0 & {\cv f_{\bar{c}_L b_R}} \\ 
0	& 0 & 0 
\end{matrix}\right)\,,
\quad
\mathcal{Y}_2^u = \left( \begin{matrix}
	0	& 0 & 0 \\ 
	0	& 0 & 0 \\ 
	0	& 0 & {\cfg g_{\bar{b}_L t_R}} 
\end{matrix}\right)\,, \nonumber
\end{eqnarray}
\begin{eqnarray}
\mathcal{Y}_1^\nu 
	=
	\left( \begin{matrix}
	0	& 0 & 0 \\ 
	0	& {\cs f_{\bar{\nu}_L \nu_R}} & 0 \\ 
	0	& 0 & {\cfg f'_{\bar{\nu}_L \nu_R}} 
\end{matrix}\right)\,,
\quad
\mathcal{Y}_1^\ell 
	=
	\left( \begin{matrix}
	0	& 0 & 0 \\ 
	0	& 0 & 0 \\ 
	0	& 0 & {\cfg f_{\bar{\tau}_L \tau_R}} 
\end{matrix}\right)\,,
\quad
\mathcal{Y}_2^\nu =\left( \begin{matrix}
	0	& 0 & 0 \\ 
	0	& {\cs g_{\bar{\nu}_L \nu_R}} & 0 \\ 
	0	& 0 & {\cfg g'_{\bar{\nu}_L \nu_R} }
\end{matrix}\right)\,,
\quad
\mathcal{Y}_2^\ell =\left( \begin{matrix}
	0	& 0 & 0 \\ 
	0	& 0 & 0 \\ 
	0	& 0 & {\cfg g_{\bar{\nu}_L\tau_R}} 
\end{matrix}\right)\,.
\end{eqnarray}
In the above equations, as well as throughout the rest of the paper, the symbol $f$ denotes the coupling of the $H_1$ doublet with the fields indicated by the subscript. Analogously, we indicate with $g$ the couplings of $H_2$. The elements of the new Yukawa matrices rendered in {\cv violet} regulate the contribution to the pseudoscalar Wilson coefficient sourced by the first diagram in Fig.~\ref{fig:S4}. The elements in {\cfg teal} enter, instead, the vector and psudovector Wilson coefficient through the remaining loop diagrams. Notice that $f_{\bar{\tau}_L\tau_R}$ is involved in both contributions, depending on the considered component of $H_1$. Lastly, terms in {\cs orange} affect exclusively the computation of the $\rksb$ anomaly presented in Sec.~\ref{sec:WCRK}. For sake of simplicity we assume real couplings and set to zero the remaining elements of the Yukawa matrices, with the understanding that their values are negligible within our effective description.

By integrating out the degrees of freedom above the $b$-quark mass scale and matching the Lagrangian in eq.~\eqref{SU2} to the effective one in eq.~\eqref{effL}, we identify the following tree-level contributions to the scalar and pseudoscalar Wilson coefficient:
\begin{eqnarray}
\label{cc3}
&&C^\tau_{SL} = -\frac{2\,m_W^2}{V_{cb}g_w^2 m_{H_1^-}^2} f_{\bar{\nu}_L\tau_R} \left( f_{\bar{b}_L c_R} - f_{\bar{c}_L b_R} \right)\,, \\
\label{cc32}
&&C^\tau_{PL} = -\frac{2\,m_W^2}{V_{cb}g_w^2 m_{H_1^-}^2} f_{\bar{\nu}_L\tau_R} \left( f_{\bar{b}_L c_R} + f_{\bar{c}_L b_R} \right)\,, 
\end{eqnarray}
where $g_w$ is the coupling constant of SM weak interactions. Notice that gauge invariance imposes $ f_{\bar{\nu}_L\tau_R} \equiv  f_{\bar{\tau}_L\tau_R}$.  Eqs.~\eqref{cc3} and~\eqref{cc32} make clear the choice of the consider Yukawa pattern, which allows $f_{\bar{b}_L c_R}$ and $f_{\bar{c}_L b_R}$ to separately source the tree-level contribution. In this way, two independent degrees of freedom regulate $C_{SL}$ and $C_{PL}$, making it possible to exploit a CKM enhancement to maintain perturbative values of the involved couplings. 

The contributions to the vector and pseudovector Wilson coefficient due to the loop diagram in Fig.~\ref{fig:S4} amount instead to
\begin{eqnarray}
\label{cc1}
&&C^{\tau\,(1)}_{VL} = -C^{\tau\,(1)}_{AL} = \left(-\frac{m_W^2}{8\pi^2 V_{cb} g_w^2}\right)\left(f_{\bar{c}_L t_R} g_{\bar{b}_L t_R} f'_{\bar{\nu}_L \nu_R} g'_{\bar{\nu}_L\nu_R }  \right) D_{dd00}[m_{\nu_R}^2, m_t^2,m_{H_1^0}^2,m_{H_2^-}^2] \,,
\\
\label{cc2}
&&C^{\tau\,(2)}_{VL} = -C^{\tau\,(2)}_{AL} =  \left(-\frac{m_W^2}{8\pi^2 V_{cb} g_w^2}\right) \left(f_{\bar{c}_L t_R} g_{\bar{b}_L t_R} f_{\bar\tau_L \tau_R} g_{\bar{\nu}_L\tau_R } \right) D_{dd00}[m_t^2,m_{H_1^0}^2,m_{H_2^-}^2, m_\tau^2]\,,
\end{eqnarray}
where we used $g_{\bar{\tau}_L\nu_R} = g'_{\bar{\nu}_L\nu_R}$ and indicated with $D_{dd00}$ the 4-point loop integral: 
\begin{eqnarray}
\label{PaVe}
&& D_{dd00}[m_{1}^2, m_2^2,m_{3}^2,m_{4}^2] = \frac{\left(2\pi\mu \right)^{4-D}}{4i\pi^2}\int d^D q \frac{q^2}{\left(q^2-m_1^2\right)\left(q^2-m_2^2\right)\left(q^2-m_3^2\right)\left(q^2-m_4^2\right)}  \,.
\end{eqnarray}
Clearly, the final expressions for the Wilson coefficients are obtained as $C^{\tau}_{VL} = C^{\tau\,(1)}_{VL} + C^{\tau\,(2)}_{VL}$ and $C^{\tau}_{AL} = C^{\tau\,(1)}_{AL} + C^{\tau\,(2)}_{AL}$.

Before moving to the discussion of the $\rksb$ anomaly, we address below the main experimental bounds that the proposed solution for $\rdsb$ faces. 

\subsection{Main experimental constraints} 
\label{sub:Bsg}

The main experimental bounds that oppose to the proposed solution for the $\rdsb$ anomaly are due to measurements of $B \rightarrow X_s \gamma$, $B^+ \rightarrow K^+ \nu \bar \nu$, and $B^0 \rightarrow K^{*0} \nu \bar \nu$. 

The introduction of extra $SU(2)$ doublets, through their Yukawa interactions, affects the dimension-5 photon and gluon dipole operators
\begin{eqnarray}
&&P_7=\frac{e}{16\pi^2}m_b\left(\bar{s} \sigma^{\mu \nu} P_R b\right) F_{\mu \nu} \,, \nn \\
&&P_8=\frac{g_3}{16\pi^2}m_b\left(\bar{s} \sigma^{\mu \nu}\,T^a P_R b\right) G^a_{\mu \nu}\,,
\end{eqnarray}
and modifies via the effective Lagrangian ~\cite{Bobeth:1999ww}
\begin{eqnarray}
\mathcal{L}_{eff} = \frac{4 G_F}{\sqrt{2}}\,V_{ts}{}^*\,V_{tb} \left(C_7 P_7 + C_8 P_8\right)\,,
\label{EffBsG}
\end{eqnarray}
the SM prediction for $B \rightarrow X_s \gamma$~\cite{Bobeth:1999ww,Hu:2016gpe,Misiak:2015xwa}. 

\begin{figure}[h]
	\centering
	\includegraphics[scale=0.8]{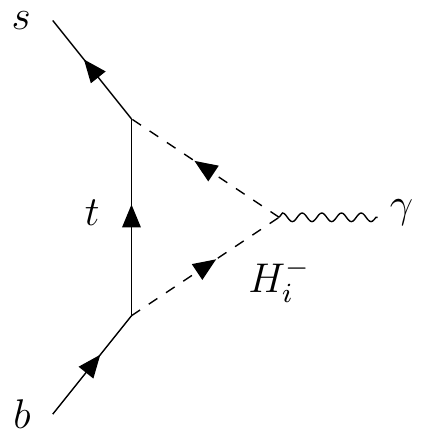}\hspace{.5cm}
	\includegraphics[scale=0.8]{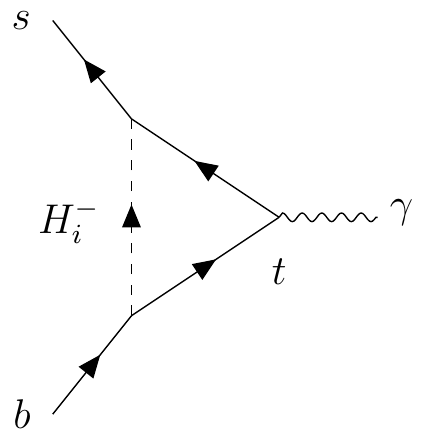}
	\caption{3HDM contributions to $b \rightarrow s \gamma$. Notice that the diagrams involve one Higgs doublet at the time.}
	\label{fig:bsg1}
\end{figure}

\begin{figure}[h]
	\centering
	\includegraphics[scale=0.38]{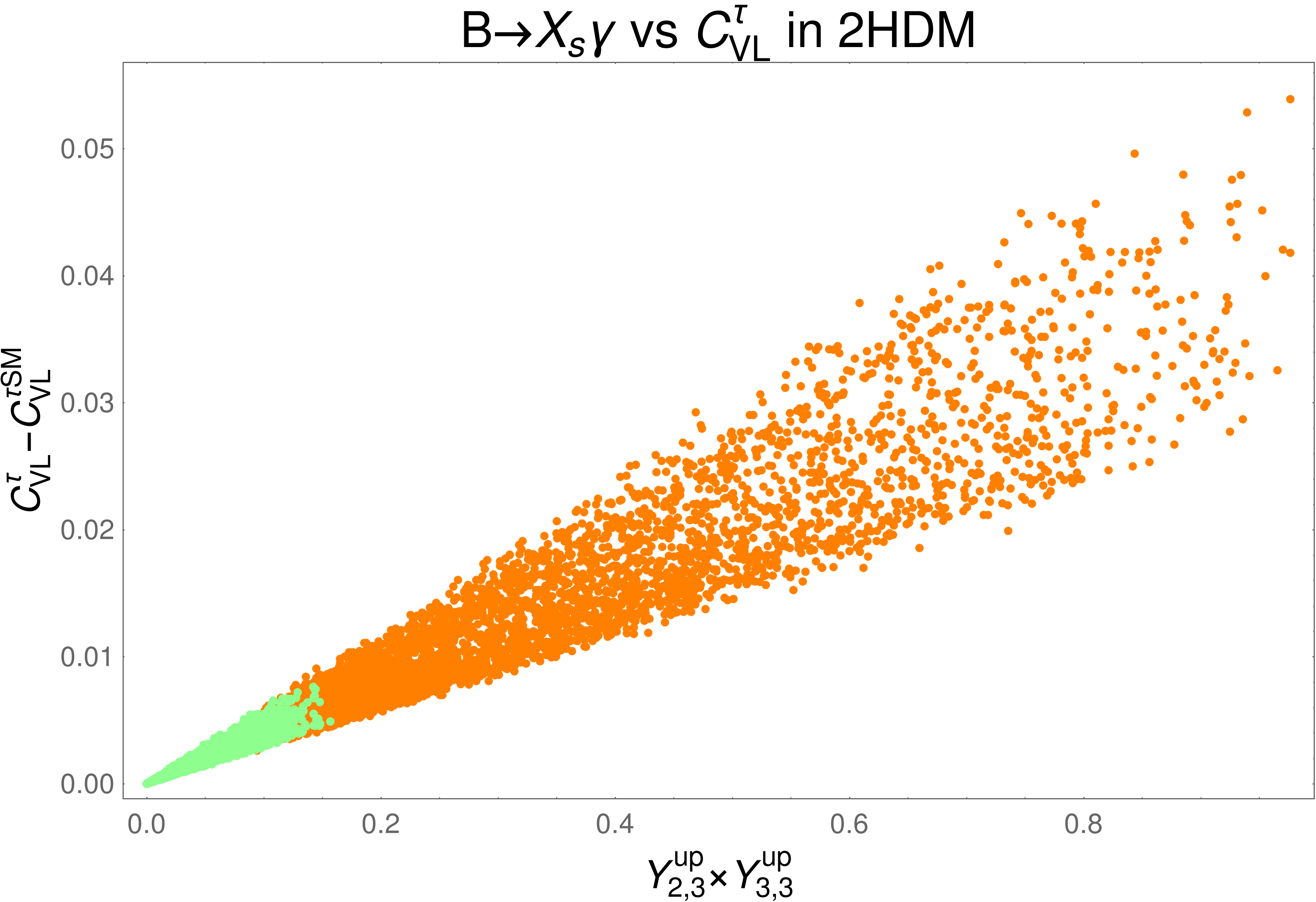}
	\caption{The contribution of the model to $C_{VL}^{\tau}$ in the 2HDM limit, where either of the new scalar doublets does not participate in the dynamics of the anomaly. The green and orange points are respectively allowed and excluded by they present measurements of $B \rightarrow X_s \gamma$. The range of masses and couplings adopted in the scan for the active extra Higgs doublet is the same as in the full 3HDM framework.}
	\label{fig:bsg2}
\end{figure}

It is then clear that measurements of this quantity limit possible new  physics contributions resulting from eq.~\eqref{yuktex}. Employing comparable couplings for both the extra doublets to induce an enhancement of the box contribution to $\rdsb$ of about a factor of $4$, in particular, generates unacceptable large contributions to $B \rightarrow X_s \gamma$ from the diagrams in Fig.~\ref{fig:bsg1}. This is manifest even in the limit where one of the two scalar doublets does not participate in the dynamics of the anomaly (2HDM limit\footnote{In this case we implicitly extend the Yukawa couplings of the active doublet to include the interactions ascribed to the excluded field within the full 3HDM.}), resulting in the bound on new contribution to $C^{\tau}_{VL}$ from $B \rightarrow X_s \gamma$ plotted in Fig.~\ref{fig:bsg2}. 

The bound therefore provides a clear indication in favour of our mechanism, which relies on separate gauge multiplets with complementary roles that generate the required amount of lepton flavour universality violation without inducing large corrections to the photon and $Z$ vertices. We remark that, within the full model, it is in principle possible to adjust the value of $\mathcal{Y}_1^d{}_{3,3}$ to cancel the new large corrections to $b \rightarrow s \gamma$ without suppressing, at the same time, the box diagrams controlling the anomaly. However, we do not pursue such a possibility because of the amount of fine tuning implied.   

\begin{figure}[h]
	\centering
	\includegraphics[scale=0.8]{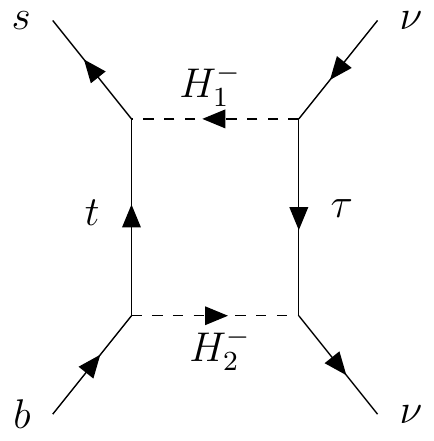}
	\caption{The main 3HDM contribution to $b \rightarrow s \nu \bar \nu$.}
	\label{fig:bkvv}
\end{figure}

As we have seen, measurements of $B \rightarrow X_s \gamma$ yield important constraints on the Yukawa couplings of the new Higgs doublet within the quark sector. In order to investigate similar constraints that potentially target the leptonic sector, we turn now our attention to $b \rightarrow s \nu_L \bar \nu_L$. Our 3HDM indeed produces a similar enhancement in the related processes via dimension-6 four-fermion operators 
\begin{eqnarray}
O_{L/R}^i = \frac{e^2}{8 \pi^2} \left(\bar s \gamma_{\mu} P_{L/R} b\right)\,\left(\bar \nu^i \gamma^{\mu} P_L \nu^i\right)\,,
\label{BsvvOp}
\end{eqnarray}
encoded in the effective Lagrangian
\begin{eqnarray}
\mathcal{L}_{eff} = \frac{4 G_F}{\sqrt{2}}\,V_{ts}{}^*\,V_{tb} \sum_{i = e,\mu \tau}\left(C^i_L O^i_L + C^i_R O^i_R\right)\,.
\label{EffBsvv}
\end{eqnarray}
The main contribution is due to the diagram in Fig~\ref{fig:bkvv}, which leads to a new flavor violating deformation of $C_L^{\tau}$ given by: 
\begin{eqnarray}
C_L^{\tau} = -\left(\frac{m_W^2 V_{cs}}{ e^2  g_w^2 V_{ts}^*}\right)\left(f_{\bar{c}_L t_R} g_{\bar{b}_L t_R} f_{\bar{\tau}_L \tau_R} g_{\bar{\nu}_L\tau_R }  \right) D_{dd00}[m_{H_1^-}^1,m_{H_2^-}^2, m_t^2] \,.
\label{bvvOpCL}
\end{eqnarray}
Notice that the absence of RHNs in the final states prevents $f'_{\bar{\nu}_L \nu_R}$ and $g'_{\bar{\nu}_L\nu_R}$ from bearing effects on the dynamics of $b \rightarrow s \nu \bar \nu$. The current upper bounds on the branching ratios are
\begin{eqnarray}
\mathcal{B}_{\mathcal{K}} = BR(B^+ \rightarrow K^+ \nu \bar \nu) < 1.7 \times 10^{-5} \, , \,\, \text{\cite{Lees:2013kla}} \nonumber \\
\mathcal{B}_{\mathcal{K}^*} = BR(B^+ \rightarrow K^{*+} \nu \bar \nu) < 4.0 \times 10^{-5} \, , \,\, \text{\cite{Lutz:2013ftz}} 
\label{boundsvv}
\end{eqnarray}
and supported by the SM results 
\begin{eqnarray}
\mathcal{B}^{SM}_{\mathcal{K}} = BR(B^+ \rightarrow K^+ \nu \bar \nu) = (3.98 \pm 0.43 \pm 0.19) \times 10^{-6}  \nonumber \\
\mathcal{B}^{SM}_{\mathcal{K}^*} = BR(B^+ \rightarrow K^{*+} \nu \bar \nu) = (9.19 \pm 0.86 \pm 0.50) \times 10^{-6} \label{SMvv}
\end{eqnarray}
have been used in Ref.~\cite{Buras:2014fpa} to derive the following $90\%$ C.L. upper bounds 
\begin{eqnarray}
\frac{\mathcal{B}_{\mathcal{K}}}{\mathcal{B}^{SM}_{\mathcal{K}}}\, < 4.3 \, ,\quad \frac{\mathcal{B}_{\mathcal{K}^*}}{\mathcal{B}^{SM}_{\mathcal{K}^*}} < 4.4 \, ,.
\end{eqnarray}

We consider the impact of these constraints in the numerical analysis presented in Sec.~\ref{sec:Discussion}. 

\section{The  $\rksb$ anomaly} 
\label{sec:WCRK}
We now turn our attention to the $\rksb$ anomaly, discussing its experimental status and showing how it can be addressed in the present framework.

\subsection{Experimental status and effective Lagrangian} 
\label{sub:Experimental status}

Another longstanding anomaly highlighted by $B$ physics experiments concerns the neutral current transition $b \rightarrow s \ell^+\ell^-$. More in detail, the LHC$b$ experiment has found anomalous values of 
\begin{equation} 
R_K = \frac{\mathcal{B}(B^+ \to K^+ \mu^+\mu^-)}{
\mathcal{B}(B^+ \to K^+ e^+e^-)}\,,
\end{equation}
and 
\begin{equation} 
R_{K^*} = \frac{\mathcal{B}(B^0 \to K^{*0} \mu^+\mu^-)}{
\mathcal{B}(B^0 \to K^{*0} e^+e^-)}\,,
\end{equation}
reporting, respectively~\cite{Aaij:2014ora, LHCB},
\begin{equation} 
R_K^\text{exp} = 0.745^{+0.090}_{-0.074} \pm 0.036 \quad \text{for } 1\,\text{GeV}^2 \le q^2 \le 6\,\text{GeV}^2\,,
\end{equation}
and
\begin{equation}
R_{K^*}^\text{exp} =
\begin{cases} 
0.66^{+0.11}_{-0.07} \pm 0.03 & \text{for } 0.045\,\text{GeV}^2 \le q^2 \le 1.1\,\text{GeV}^2\,,\\
0.69^{+0.11}_{-0.07} \pm 0.05 & \text{for } 1.1\,\text{GeV}^2 \le q^2 \le 6\,\text{GeV}^2\,,\\
\end{cases}
\end{equation}
where $q^2$ is the invariant mass of the final state di-lepton system.

The corresponding SM predictions\footnote{As computed with the \texttt{flavio-0.21.2} package~\cite{Altmannshofer:2017fio}.}, suppressed by the GIM mechanism, amount to 
\begin{equation} 
R_K^\text{SM} = 1.0004 \pm 0.0002\,,
\end{equation}
and
\begin{equation}
R_{K^*}^\text{SM} =
\begin{cases} 
0.926 \pm 0.003\,,\\
0.9965 \pm 0.0005\,,\\
\end{cases}
\end{equation}
giving consequently rise to a discrepancy with a significance of about 5$\sigma$~\cite{Altmannshofer:2017fio} depending on the details of the fit.

The low-energy effective Lagrangian describing the $b \rightarrow s \ell \ell$ transition is 
\begin{eqnarray} \label{eq:LagBs}
- \mathcal{L}_{bs} = -\frac{4\,G_F}{\sqrt{2}}\,V_{tb}V^*_{ts}\frac{e^2}{16\,\pi^2}\sum_i \left(\mathcal{C}_i\mathcal{O}_i + \mathcal{C}'_i\mathcal{O}'_i\right)\,,
\end{eqnarray}
where the dimension-6 operators $\mathcal{O}_i$ are defined as 
\begin{eqnarray}
&\mathcal{O}_7 = (\bar{s}\,P_L\,b)(\bar{l} l)\,,&  \quad \mathcal{O}'_7 = (\bar{s}\,P_R\,b)(\bar{l} l) \, , \nn \\
& \mathcal{O}_{8} = (\bar{s}\,P_L\,b)(\bar{l}\gamma_5 l)\,,&   \quad \mathcal{O}'_{8} = (\bar{s}\,P_R\,b)(\bar{l}\gamma_5 l) \, , \nn \\
& \mathcal{O}_9 = (\bar{s}\,\gamma_{\mu}P_L\,b)(\bar{l}\gamma^{\mu} l)\,,&  \quad \mathcal{O}'_9 = (\bar{s}\,\gamma_{\mu}P_R\,b)(\bar{l}\gamma^{\mu} l) \, , \nn \\
& \mathcal{O}_{10} = (\bar{s}\,\gamma_{\mu}P_L\,b)(\bar{l}\gamma^{\mu}\gamma_5 l)\,,& \quad \mathcal{O}'_{10} = (\bar{s}\,\gamma_{\mu}P_R\,b)(\bar{l}\gamma^{\mu}\gamma_5 l) \, .
\end{eqnarray}
Global fits of the anomalies presently converge on a preferred sets of Wilson 
coefficients~\cite{Altmannshofer:2017fio,Descotes-Genon:2013wba,Descotes-Genon:2015uva,DiChiara:2017cjq}, with the highest pull given by $C_9-C_9^{SM} \simeq -1.21$ or $C_9-C_9^{SM} = -\left(C_{10}-C_{10}^{SM}\right) \simeq -0.67$ that sets an almost 5$\sigma$ discrepancy with respect to the SM predictions. 

\subsection{Explaining $\rksb$ in the 3HDM} 
\label{sec:boundsRK}

The explanation of the $\rksb$ anomaly in our 3HDM reflects, for the most part, the steps required to account for $\rdsb$ one. The possible role of the scalar operators is once again strongly constrained, in this case by observations of $B_s \rightarrow \mu^+ \mu^-$. We therefore seek an enhancement of the $V-A$ quark current, along with a suppression of scalar Wilson coefficient, in order to fit the current measurements. 

The explanations of the two anomalies share a common origin in the couplings of the quark sector $f_{\bar{c}_Lt_R}$ and $g_{\bar{b}_Lt_R}$. The additional source of lepton flavor universality violation is here implemented through the interactions with the RHNs and muons
\begin{eqnarray}
- \mathcal{L} &\supset&  f_{\bar{\mu}_L\nu_R} H_1^- \bar{\mu}_L\nu_R +  g_{\bar{\mu}_L\nu_R} H_2^- \bar{\mu}_L\nu_R +  h.c.  \, ,
\label{LagNPRK}
\end{eqnarray}
where $f_{\bar{\mu}_L\nu_R} \equiv f_{\bar{\nu}_L \nu_R}$, $ g_{\bar{\mu}_L\nu_R} \equiv g_{\bar{\nu}_L \nu_R}$. 

\begin{figure}[H]
	\centering
	\includegraphics[width=.25\linewidth]{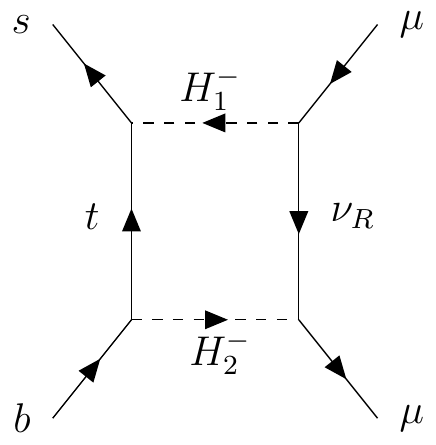}
	\caption{The diagram responsible for reproducing the $\rksb$ anomaly within the considered 3HDM.}
	\label{fig:S7}
\end{figure}

Our choice of couplings for the charged scalar fields results in vectorial Wilson coefficients that obey $C_9 = -C_{10}$, with the contribution of the diagram in Fig.~\ref{fig:S7} amounting to
\begin{eqnarray}
- C_9 = C_{10} = -\frac{m^2_W}{4\pi\,\alpha g_w^2} \frac{V^*_{cs}}{V^*_{ts} } \left(f_{\bar{c}_Lt_R} g_{\bar{b}_Lt_R}  f_{\bar{\mu}_L\nu_R} g_{\bar{\mu}_L\nu_R} \right) D_{dd00}[m_{t}^2, m_{H_1^-}^2, m_{H_2^-}^2]\, ,
\end{eqnarray}
where $D_{dd00}$ is the same 4-points scalar integral given in eq.~\eqref{PaVe}. Notice that the $t_R \bar{s}_L$ coupling is obtained from the $t_R \bar{c}_L$ one via the CKM element $V^*_{cs}$.

\section{Results}
\label{sec:Discussion}

We gather here the results obtained for the $\rdsb$ and $\rksb$ anomalies through the analysis detailed in the previous sections, discuss their compatibility within the present framework and remark on the possible impact of new determinations of $\epsilon'/\epsilon$ on our conclusions.

\subsection{Numerical analysis}

As for the anomalies, the joint effect of the tree-level scalar contribution and of the operator $\left[\bar{c} \gamma^{\mu} P_L b\right]\left[\bar{\ell}\gamma_{\mu} P_L \nu_{\ell}\right]$, generated by the same particles at the one-loop level, allows access to the $2\sigma$ and $1\sigma$ $\rdsb$ regions for values of the Yukawa sector well within the perturbative regime. Remarkably, once supplemented with an extra coupling to muons, we find that the same interactions between quarks and the new scalar doublets allow also to explain the $\rksb$ signal. 

\begin{figure}[htbp]
	\centering
	\includegraphics[width=.42\linewidth]{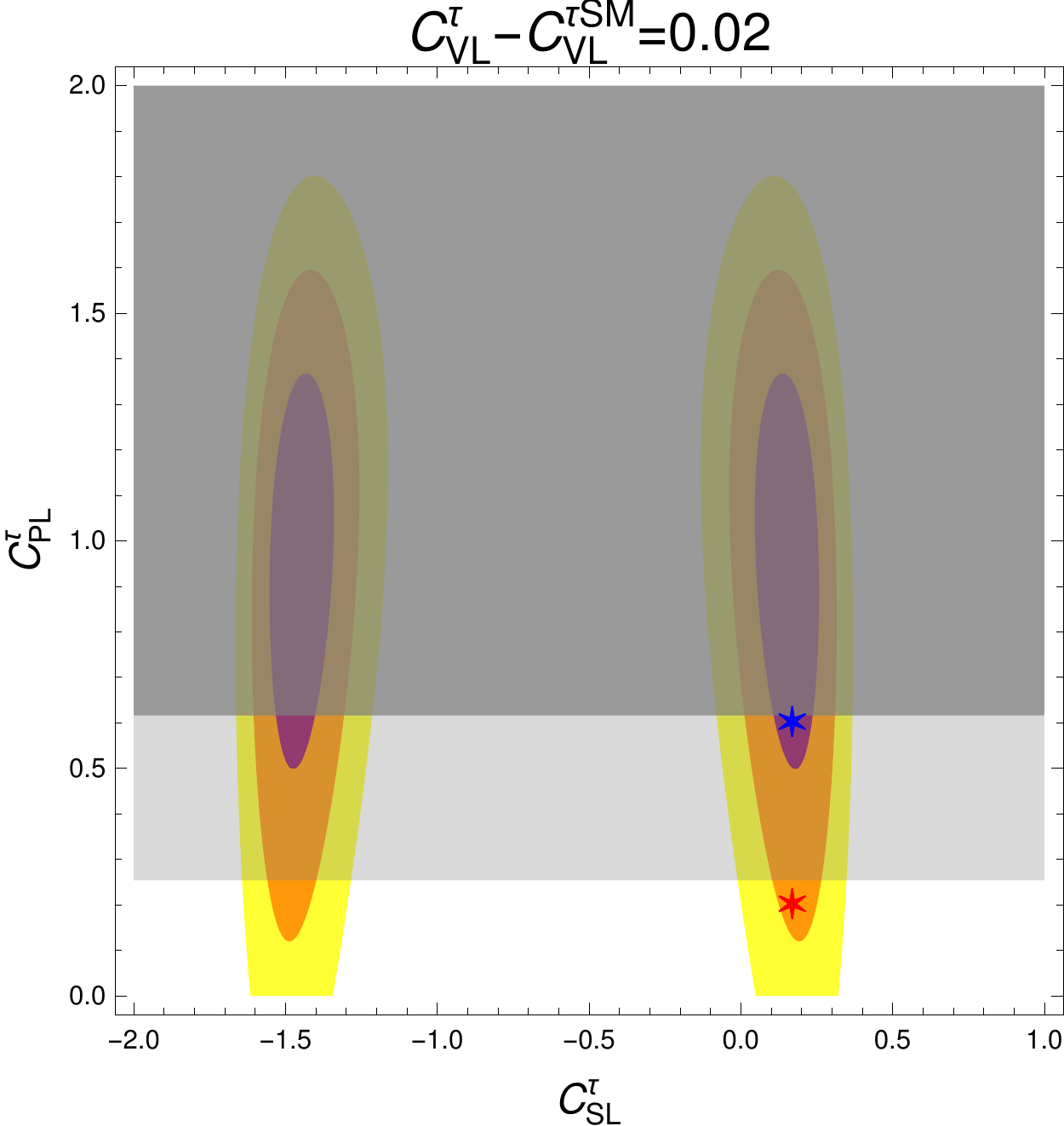}
	\hspace{.5cm}
	\includegraphics[width=.42\linewidth]{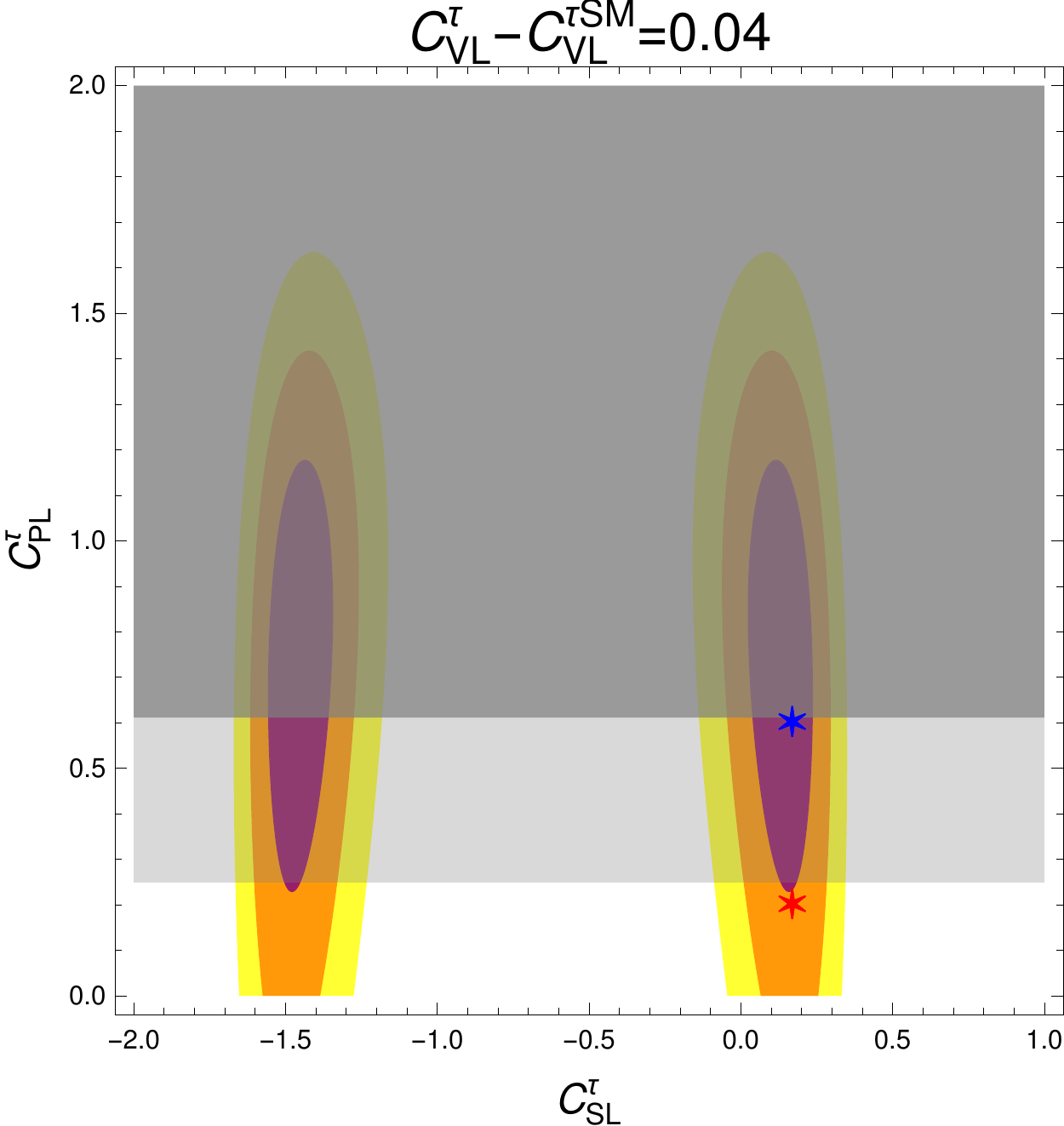}
	\caption{The $\rdsb$ anomaly: benchmark points obtained with the considered 3HDM. The red, orange and yellow areas respectively indicate the 68\%, 95\% and 99\% confidence intervals selected by the joint fit of $\rd$ and $\rds$. The shaded light (dark) gray areas represent instead the current bound from the measured $B_c$ lifetime assuming a 10\% (30\%) maximal allowed deviation. The two panels differ by the indicated value of the vector Wilson coefficients.}
	\label{fig2}
\end{figure}

The result concerning the $\rdsb$ anomaly is illustrated in isolation in Fig~\ref{fig2}. Here we show four benchmark points in the space of Wilson coefficients obtained through the contributions detailed in Sec.~\ref{sec:WC}. The points in red comply with the $10\%$ lifetime bounds on $B_c$ decay, while the blue ones refer to the corresponding $30\%$ limit. The areas shaded in red, orange and yellow indicate the $68\%$, $95\%$ and $99\%$ confidence interval indicated by current measurements, respectively. The two panels differs by the considered value of the vector Wilson coefficients $C^{\tau}_{VL} = - C^{\tau}_{AL}$, induced by the 3HDM at the one-loop level. As shown in the right panel, larger values of this quantity allow to fit the anomaly with a contribution from $C^{\tau}_{PL}$ small enough to comply with the $B_c$ lifetime bounds.

\begin{figure}[h]
	\centering
	\includegraphics[scale=0.5]{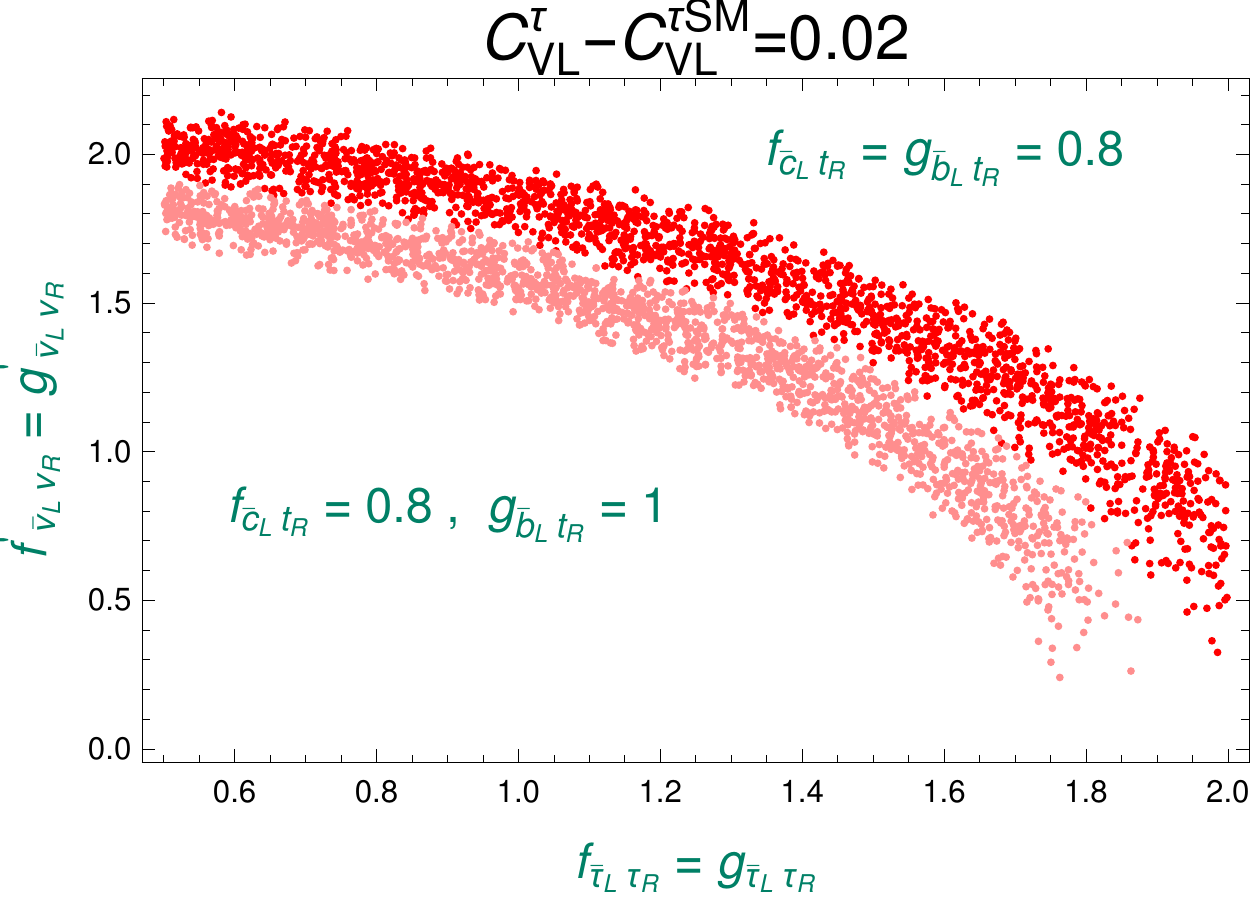}
	\includegraphics[scale=0.5]{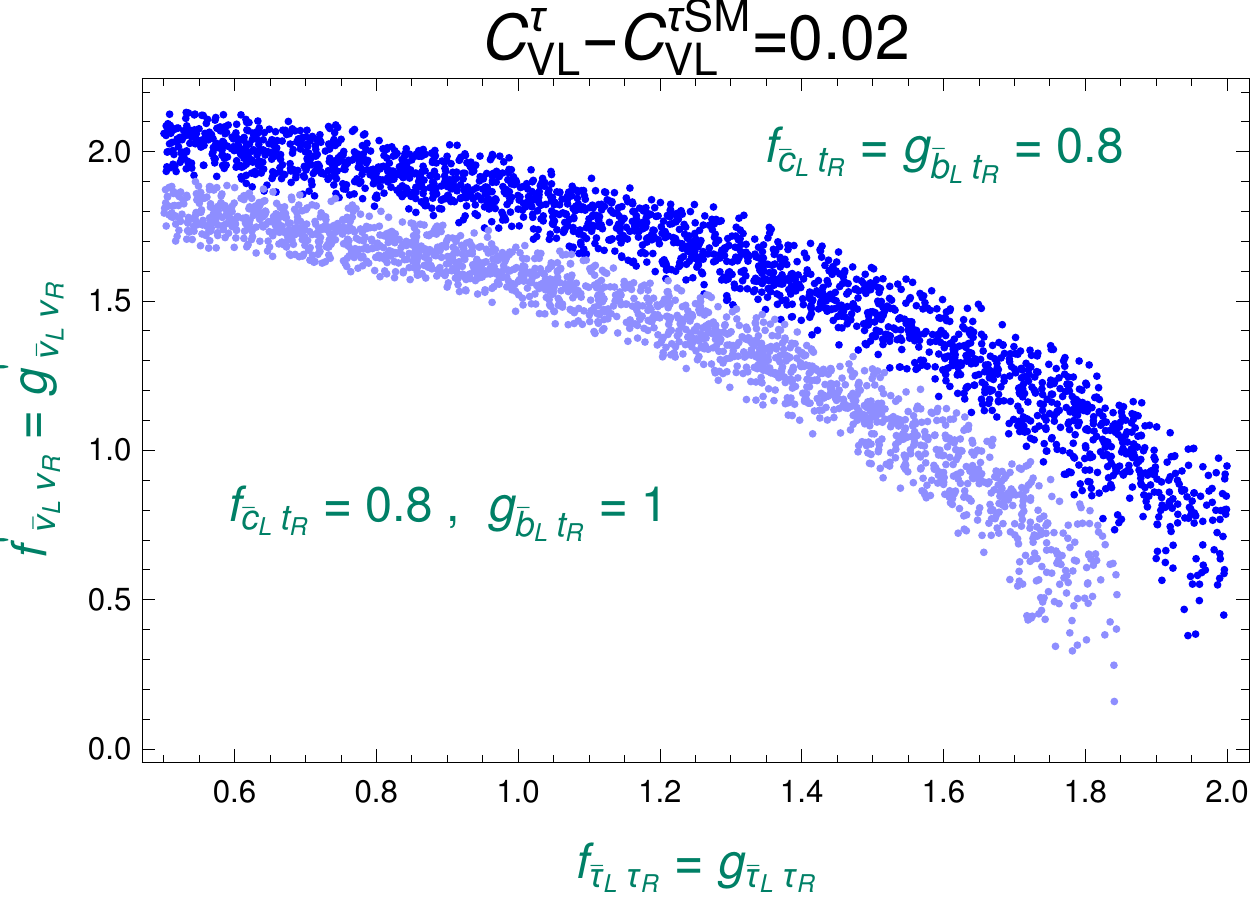}\\
	\includegraphics[scale=0.5]{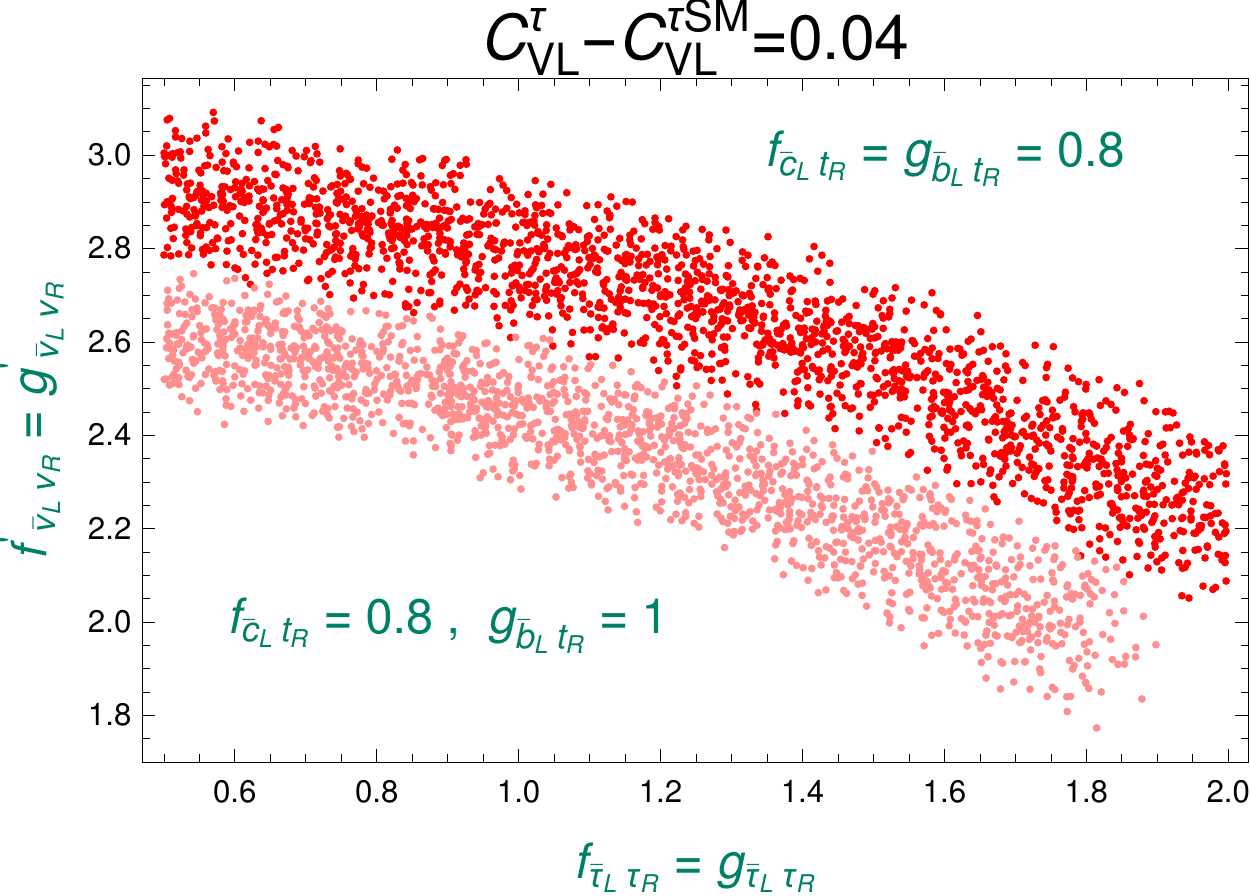}
	\includegraphics[scale=0.5]{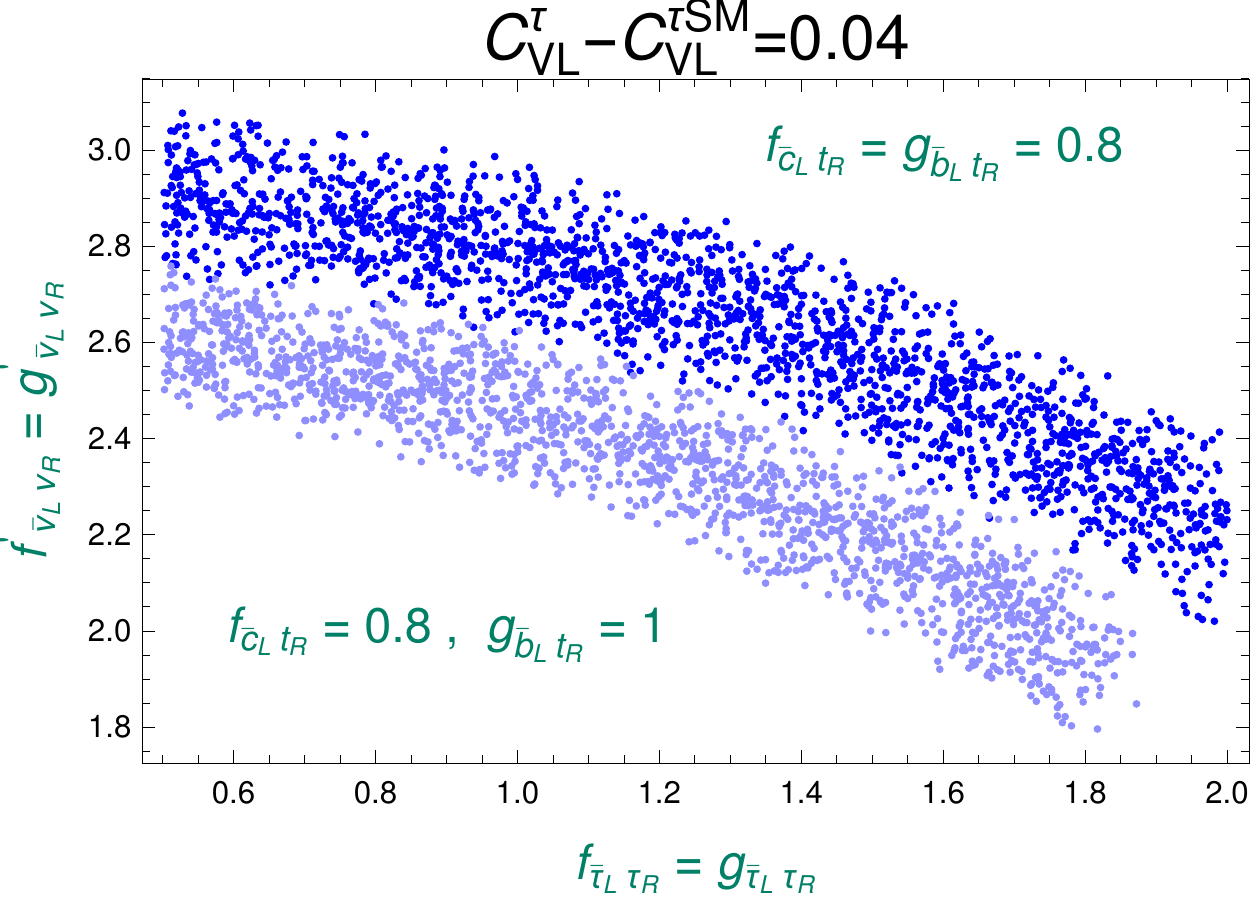}
	\caption{Values of the lepton and RHN Yukawa couplings resulting in the benchmark points presented in Fig.~\ref{fig2} for the indicated choice of the quark Yukawa couplings. All points respect the bounds discussed in Sec.~\ref{sub:Bsg}.}
	\label{fig3}
\end{figure}

The benchmark points have been obtained by maximizing the one-loop contributions encoded in the vector and pseudovector operators, which depend on the set of couplings rendered in teal in eq.~\eqref{yuktex}. Considering the collider phenomenology analysis presented in Ref.~\cite{Fraser:2018aqj}, we set the values of the new quark Yukawa couplings to ${\cfg f_{\bar{c}_L t_R}} = 0.8$ and ${\cfg g_{\bar{b}_L t_R}} = 0.8$ or ${\cfg g_{\bar{b}_L t_R}} = 1$. The magnitude of the RHN and lepton couplings ${\cfg f'_{\bar{\nu}_L \nu_R}} = {\cfg g'_{\bar{\nu}_L \nu_R} }$ and ${\cfg f_{\bar{\tau}_L \tau_R}}={\cfg g_{\bar{\nu}_L\tau_R}}$, show in Fig~\ref{fig3}, is then obtained by setting the Wilson coefficients to the indicated values. 

\begin{figure}[h]
	\centering
	\includegraphics[scale=0.6]{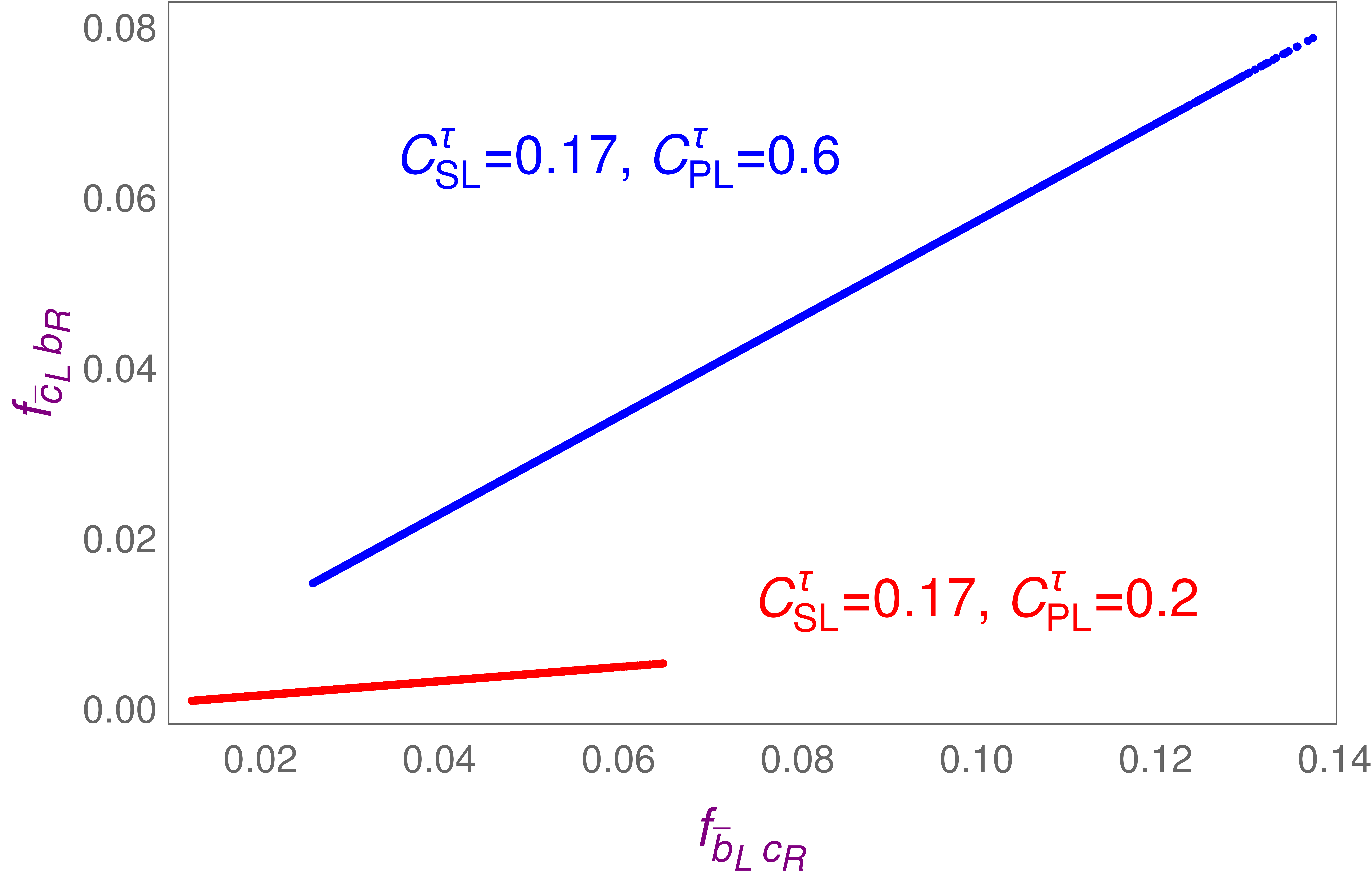}
	\caption{Values of the quark Yukawa couplings that result in the scalar and pseudoscalar Wilson coefficients indicated in Fig.~\ref{fig2}. The plot is obtained by letting the scalar masses vary in the $300-350$ GeV range and setting $ f_{\bar{\nu}_L\tau_R}$ as indicated by the vector Wilson coefficients.}
	\label{fig4}
\end{figure}

The subdominant tree-level contribution of $C^{\tau}_{SL}$ and $C^{\tau}_{PL}$ is subsequently obtained through Eq.~\ref{cc3}, in compliance with the $B_c$ lifetime bounds. The required magnitude of the involved couplings ${\cv f_{\bar{b}_L c_R}}$ and ${\cv f_{\bar{c}_L b_R}}$, which do not enter the vector contribution, is presented in Fig~\ref{fig4}. Here we let the mass of the new scalar particles vary in the selected range and set $ f_{\bar{\nu}_L\tau_R} \equiv f_{\bar{\tau}_L\tau_R}$ according to the vector operator. 

Finally, the $\rksb$ anomaly can be accommodated by setting the remaining independent parameters of the model ${\cs f_{\bar{\nu}_L \nu_R}} = {\cs g_{\bar{\nu}_L \nu_R}}$, resulting in $O_{9} = - O_{10} = -0.67$, as required by global fits of the anomalous signal.  

\FloatBarrier

\FloatBarrier

\subsection{Compatibility of the two anomalies}

\begin{figure}[htbp]
	\centering
	\includegraphics[width=.25\linewidth]{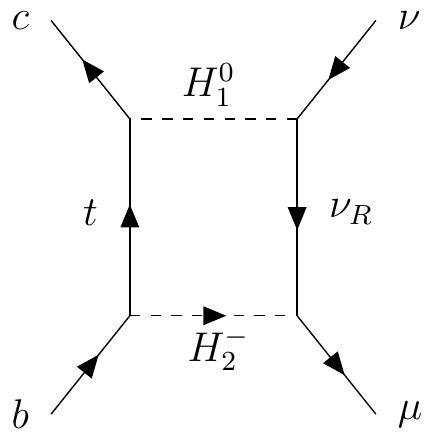}
	\hspace{.8cm}
    \includegraphics[width=.44\linewidth]{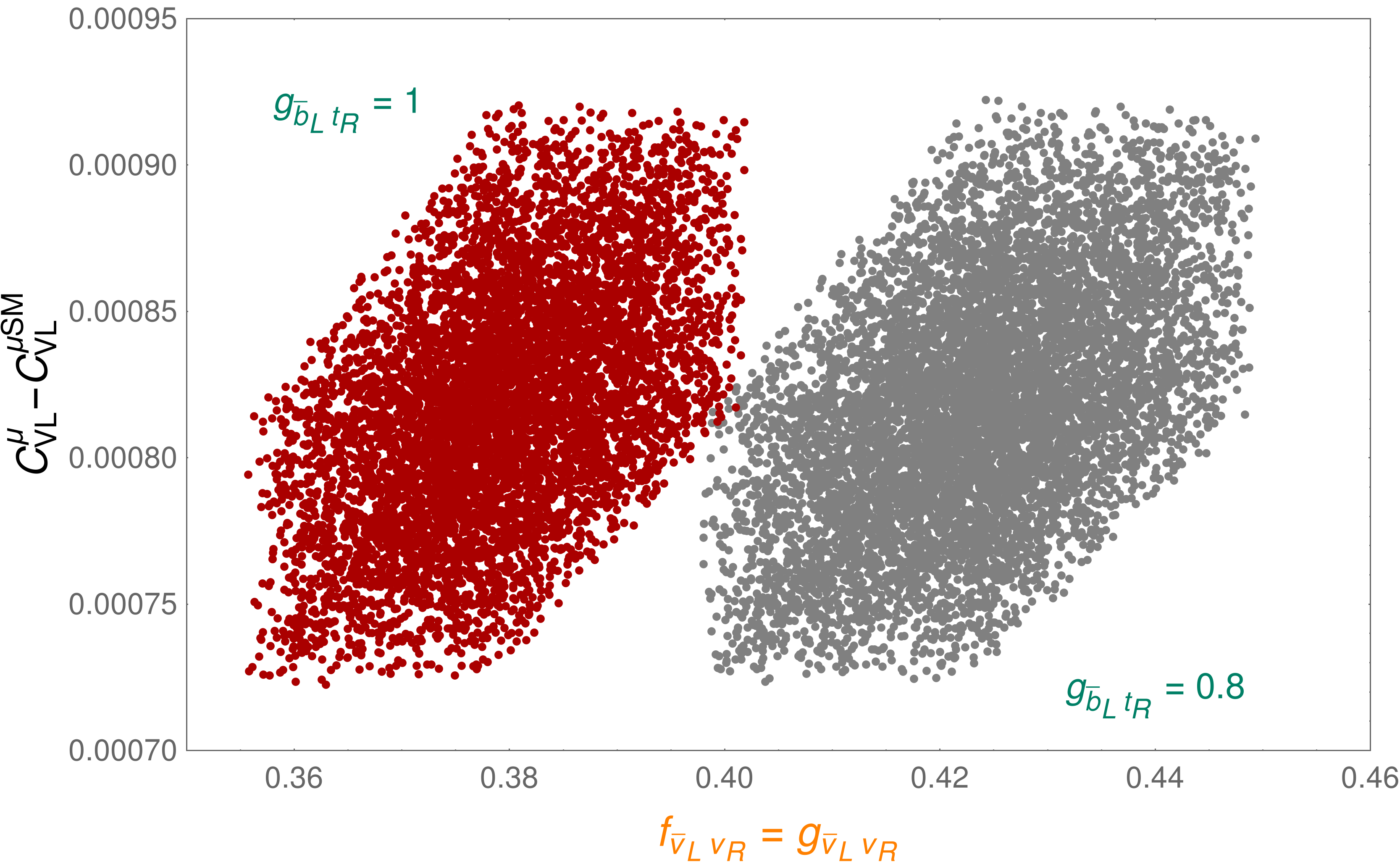}
    	\caption{Left panel: the diagram that hinders a simultaneous explanations for $\rdsb$ and $\rksb$ in the present framework. Right panel: values of $C^{\mu}_{VL}$ resulting from the Yukawa couplings used to explain the $\rksb$ anomaly.}
	\label{fig:S8}
\end{figure}

A possible problem for the simultaneous explanations of the charged and neutral current flavour anomalies arises from the definition of $\rdsb$ 
\begin{equation} 
R_{D^{(*)}} = \frac{\mathcal{B}(\bar{B} \to D^{(*)} \tau \bar{\nu})}{
	\mathcal{B}(\bar{B} \to D^{(*)} \ell \bar{\nu})} \,,
\end{equation}
where the denominator contains an average over muons and electrons. It could be consequently thought that the couplings introduced in eq.~\eqref{SU2} to explain the $\rksb$ anomaly (in orange), if sizeable, may dilute the yield of the model to $\rdsb$. However, because the contribution of the diagram in the left panel of Fig.~\ref{fig:S8} is negligible, the analyses of $\rdsb$ and $\rksb$ are essentially uncorrelated within the present framework. This is due to the fact that the the couplings employed in the muon sector to explain the $\rksb$ anomaly are much smaller than the ones entering the expression for $\rdsb$, as shown in the right panel of Fig.~\ref{fig:S8}.

\FloatBarrier

\subsection{Impact of new $\epsilon'/\epsilon$ determinations}

To conclude our analysis we investigate the robustness of our model with respect to another flavour observable that recently received increasing attention~\cite{Buras:2018ozh, Gisbert:2017vvj}: $\epsilon'/\epsilon$. 

This quantity, which quantifies the direct CP violation in $K\to \pi \pi$ decays, has been debated in the literature as a further indication of new physics in flavour measurements. The discussion has been revived after the lattice QCD result obtained by the RBC-UKQCD group~\cite{Blum:2015ywa,Bai:2015nea}
\begin{eqnarray}
Re(\epsilon'/\epsilon) = 1.38(5.15)(4.59) \times 10^{-4}\,,
\end{eqnarray}
was corroborated by a large $N_c$ dual QCD estimate~\cite{Buras:2015xba,Buras:2015yba} of
\begin{eqnarray}
Re(\epsilon'/\epsilon) = (1.9 \pm 4.5) \times 10^{-4}\,.
\end{eqnarray}
The mutual agreement of these independent estimates of the SM contribution results in a $2\sigma$ deviation from the current experimental measure~\cite{Batley:2002gn,AlaviHarati:2002ye,Abouzaid:2010ny}
\begin{eqnarray}
Re(\epsilon'/\epsilon)_{exp} = (16.6 \pm 2.3) \times 10^{-4}\, .
\end{eqnarray}

Presently the origin of the anomaly remains unclear as misdeterminations of the SM contribution alone could explain the mentioned deviation~\cite{Gisbert:2017vvj}.  In light of this ambiguous signal, we discuss below how new determinations of the SM contribution into $\epsilon'/\epsilon$ could constrain, or be rectified, in the present framework. 

The contribution of a charged scalar to the effective $d\rightarrow sg$ Hamiltonian 
\begin{eqnarray} \label{CMOs}
	\mathcal{H}_{d\rightarrow sg} = - \frac{G_F}{\sqrt{2}}\,V^*_{ts}V_{td}\left( C_{8 g} O_{8 g} + C'_{8 g} O'_{8 g}  \right)\, , 
	\end{eqnarray}
	where the chromomagnetic dipole operators (CMOs) are 
	\begin{eqnarray} \label{CMOs3}
	O_{8 g} = \frac{g_s}{8 \pi^2} \bar{s} \sigma^{\mu \nu} T^a P_{L} d G_{\mu \nu}^a \,, \quad
	O_{8 g} = \frac{g_s}{8 \pi^2} \bar{s} \sigma^{\mu \nu} T^a P_{R} d G_{\mu \nu}^a \,.
	\end{eqnarray}
results in a short distance modification of the SM dynamics which, without introducing further sources of CP violation, alleviates the tension with the current experimental results~\cite{Chen:2018vog,Chen:2018ytc}.  

The determination of the hadronic matrix element for $K^0 \rightarrow \pi \pi$  in dual QCD~\cite{Buras:1985yx,Bardeen:1986vp,Bardeen:1986uz,Bardeen:1986vz,Bardeen:1987vg}, allows an estimate of order of magnitude needed for the Wilson coefficients in eq.~\eqref{CMOs} to reproduce the $\epsilon'/\epsilon$ measurement~\cite{Buras:2015yba}
\begin{eqnarray} \label{DirectCP}
Re\left(\epsilon'/\epsilon\right)_{8g} \sim - \left(1.85 \times 10^5 GeV\right) \times Im\left(C^-_{8g}\right) \,\, ,
\end{eqnarray} 
where $C^-_{8g}$ is the combination 
\begin{eqnarray}
C^-_{8g} = - \frac{G_F}{\sqrt{2}}\,V^*_{ts}V_{td} \left(  m_d C'_{8 g} - m_s C_{8 g}\right)	\,.
\end{eqnarray}
We stress that eq.~\eqref{DirectCP} is obtained under the assumption of a SM-like value for the indirect CP violation in $\epsilon_K$, enforced in the framework at hand by the same bounds targeting the mass splitting and CP violation of the neutral $K$ system~\cite{Blum:2009sk}. As previously shown in Ref.~\cite{Fraser:2018aqj}, these constraints do not significantly limit the parameter space selected by the anomalous signals.  

\begin{figure}[htbp]
	\centering
	\includegraphics[width=.25\linewidth]{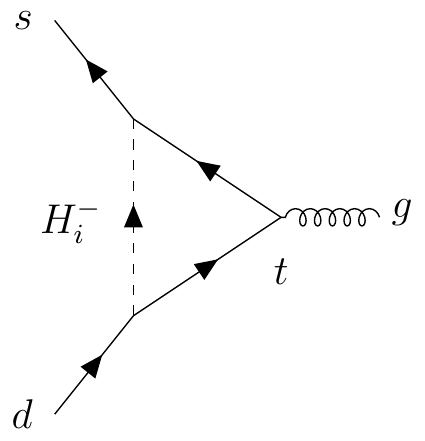}
	\hspace{.8cm}
	\includegraphics[width=.44\linewidth]{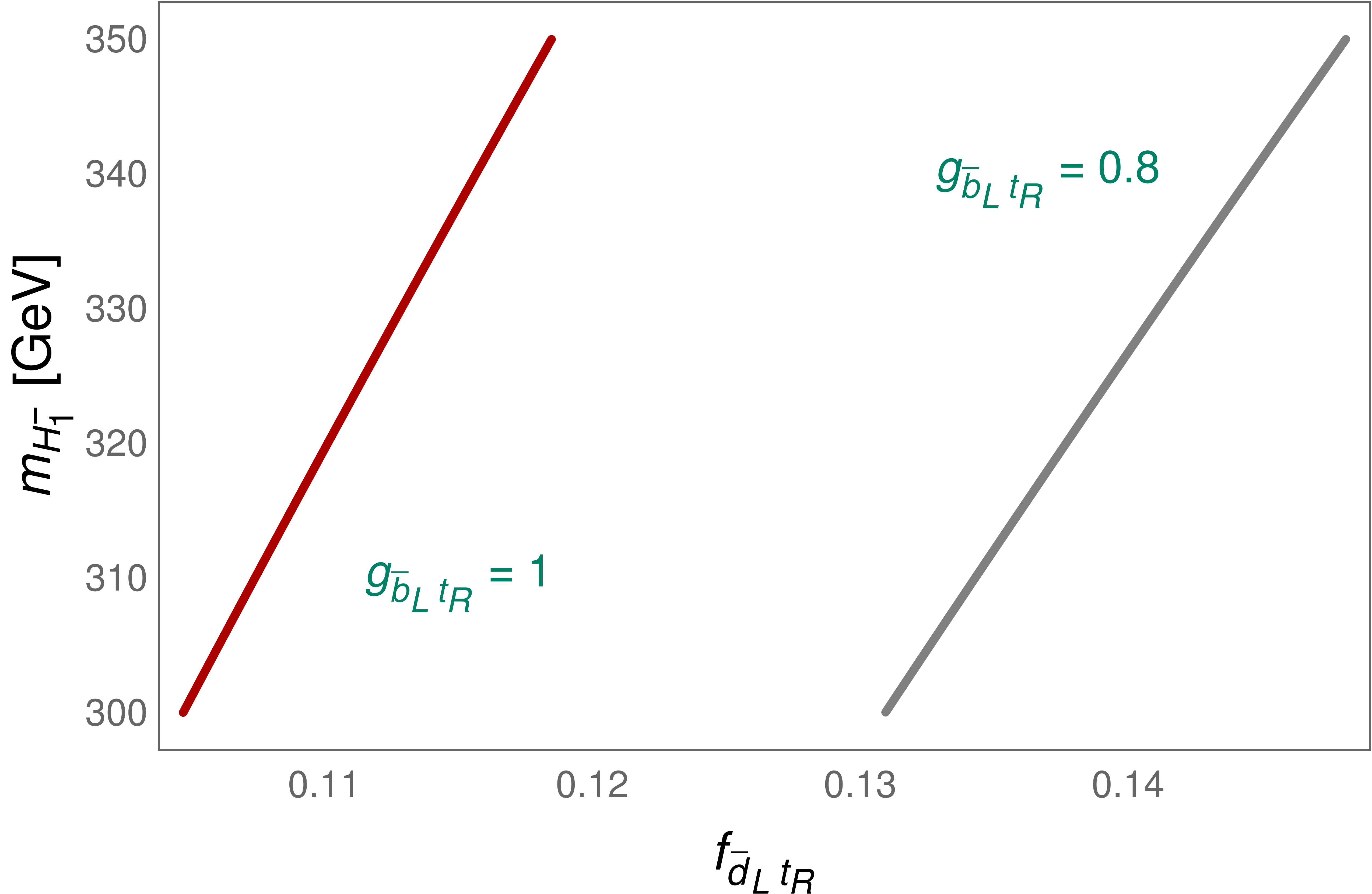}
	\caption{Left panel: The dominant 3HDM contribution to $\epsilon'/\epsilon$ . Right panel: values of ${\color{black} f_{\bar{d}_L t_R}}$ required to match the measured value $Re\left(\epsilon'/\epsilon\right)_{8g}  = 10^{-3}$ for the indicated choices of the quark Yukawa couplings.}
	\label{fig:OSg}
\end{figure} 

With the Yukawa texture in eq.~\eqref{yuktex}, the same interactions behind our explanations of the $\rdsb$ and $\rksb$ anomalies yield new contributions to  the CMOs that source $\epsilon'/\epsilon$. In particular, the propagation of $H^{-}_2$ in Fig.~\ref{fig:OSg} results in
\begin{eqnarray}
C^-_{8g}{}_{H_2} &=& \frac{1}{4} V_{tb} V^{*}_{ts} \, |{\cfg g_{\bar{b}_L t_R}}|^2 \left\{m_d \left(C_{cc12}[m_{H_2^-}^2, m_t^2, m_t^2] + C_{cc2}[m_{H_2^-}^2, m_t^2, m_t^2] + C_{cc22}[m_{H_2^-}^2, m_t^2, m_t^2]\right) + \right. \\ &&\left. - m_s \left(C_{cc1}[m_{H_2^-}^2, m_t^2, m_t^2] + C_{cc11}[m_{H_2^-}^2, m_t^2, m_t^2] + C_{cc12}[m_{H_2^-}^2, m_t^2, m_t^2]\right) \right\} \nn \\\nn &=& V_{tb} V^{*}_{ts} \,\frac{m_{d}-m_s}{48\left(m_{H_2^-}^2 - m_t^2\right)^4}  |{\cfg g_{\bar{b}_L t_R}}|^2  \left(2 m_{H_2^-}^6 + 3 m_{H_2^-}^4 m_t^2 - 6 m_{H_2^-}^2 m_t^4 - 6 m_{H_2^-}^4 m_t^2 \log\left(m_{H_2^-}^2/m_t^2\right) + m_t^6 \right) \, , 
\end{eqnarray}
and, in turn, in value of $\epsilon'/\epsilon$ comparable in magnitude to the current SM estimate. Therefore, should a revised computation of the SM contribution explain the observed discrepancy, precision measurements of this parameter will certainly offer a new way to probe our scenario.        

On the contrary, should the $\epsilon'/\epsilon$ measurement require the presence of new physics, our framework could offer a possible solution through a minimal modification of the proposed Yukawa texture   
\begin{eqnarray}
\mathcal{Y}_1^d =\left( \begin{matrix}
0	& 0 & 0 \\ 
0	& 0 & {\cv f_{\bar{c}_L b_R}} \\ 
0	& 0 & 0 
\end{matrix}\right)\, \quad \rightarrow \quad
\mathcal{Y}_1^d =\left( \begin{matrix}
0	& 0 & 0 \\ 
0	& 0 & {\cv f_{\bar{c}_L b_R}} \\ 
0	& 0 & {\color{black} f_{\bar{s}_L t_R}} 
\end{matrix}\right)\,,
\end{eqnarray}
where the new diagonal down-Yukawa coupling ${\color{black} f_{\bar{s}_L t_R}}$ results, via the $H_1^{-}$ contribution shown in Fig.~\ref{fig:OSg}, in the term
\begin{eqnarray}
C^-_{8g}{}_{H_1} &=& -\frac{m_t}{4} V_{tb} V^{*}_{cs} \left( {\cv f_{\bar{c}_L b_R}} {\color{black} f_{\bar{s}_L t_R}}\right) \left(C_{cc1}[m_{H_1^-}^2, m_t^2, m_t^2] + C_{cc2}[m_{H_1^-}^2, m_t^2, m_t^2]\right) \nn \\ &=& V_{tb} V^{*}_{cs} \,\frac{m_{t}}{8\left(m_{H_1^-}^2 - m_t^2\right)^3}  \left( {\cv f_{\bar{c}_L b_R}} {\color{black} f_{\bar{s}_L t_R}}\right) \left(3 m_{H_1^-}^4 - 4 m_{H_1^-}^2 m_t^2 - 2 m_{H_1^-}^4 \log\left(m_{H_1^-}^2/m_t^2\right) + m_t^4 \right) \, .
\end{eqnarray}
Notice the relevance of the Yukawa term ${\cfg f_{\bar{c}_L t_R}}$, which is therefore pivotal for the explanation of \textit{all of the anomaly discussed in the present work} and offers a clear signature for the investigation of our proposal.

In Fig.~\ref{fig:OSg} we assess the value of ${\color{black} f_{\bar{d}_L t_R}}$ required to have $Re\left(\epsilon'/\epsilon\right)_{8g}  = 10^{-3}$, considering the values of ${\cfg f_{\bar{c}_L t_R}}$ and ${\cfg g_{\bar{b}_L t_R}}$ used in the benchmark points of the previous section. Clearly, it is sufficient to set ${\color{black} f_{\bar{d}_L t_R}} \sim 10^{-1}$ to enhance the 3HDM contribution into the CMOs and explain the $\epsilon'/\epsilon$ anomaly on top of the $\rdsb$ and $\rksb$ signals.

\section{Discussion and conclusions} 
\label{sec:Conclusions}

We attempted to demystify the implications of the $\rdsb$ and $\rksb$ anomalies in the context of physics beyond the standard model. Although these signals seem to indicate new lepton flavour violating physics at very different scales, we demonstrated that this fact does not necessarily imply the existence of exotic new physics. In fact, the full dynamics responsible for all of the analyzed measurements can still be addressed within a more conventional framework, using model building elements already available in the standard model.

Specifically, we have shown that the present $\rdsb$ and $\rksb$ anomalies can be simultaneously explained in a 3 Higgs doublets model extended with right-handed neutrinos. The results plotted in Fig.~\ref{fig2} demonstrate that the model predicts values of $\rdsb$ within the $1\sigma$ region indicated by the experiments, in accordance with the remaining phenomenological constraints and for perturbative values of the involved Yukawa couplings. The result is achieved owing to the interplay between the loop induced vector operators and tree-level scalar operators in eq.~\eqref{effL}, which depend on different sets of parameters. We have furthermore shown that in the considered scheme the $\rdsb$ and $\rksb$ anomalies arise from independent interactions of the additional scalar doublets, and consequently found a simultaneous explanation for the two measurements.

The robustness of our results was tested against a further flavour observable, $\epsilon'/\epsilon$. Once again, we find that within the proposed 3HDM this anomaly can be modeled independently from the remaining flavour physics signals.
Should the current deviation in this parameter be confirmed as an effect of new physics, our proposal would then provide a first framework able to simultaneously explain all of the mentioned anomalies.  

A crucial aspect of our work is the inclusion of GeV scale right-handed neutrinos, which allowed for the sizable loop-level contributions needed to explain the analyzed signals. It is remarkable that the presence of these particles in Nature is advocated also in connection to other open problems of contemporary physics~\cite{Asaka:2005an,Canetti:2012kh,Eijima:2018qke}, and our framework therefore connects the $\rdsb$ and $\rksb$ flavour anomalies to the well-known phenomenology of neutrino masses and to the puzzle of the baryon asymmetry of the Universe. The considered right-handed neutrinos are also a primary target of the next generation beam-dump experiment SHIP, at CERN, which has therefore the potential to directly test our scenario on top of complementary collider searches for new scalar particles.

In the light of the proposed scheme, it is natural that the first signals of new physics would be detected in low-energy flavour experiments rather than in high-energy collider searches and precision observables: because the GeV-scale right-handed neutrinos phenomenology is intrinsically a low-energy phenomenon, the first signals of their presence naturally occur in the loop processes behind the flavour observables under discussion, preceding the potential  collider signatures of still undiscovered scalar particles.


\section*{Acknowledgement} 
This work is supported by the Estonian Research Council grants PRG356, MOBTT5, IUT23-6 and the ERDF Centre of Excellence project TK133.



\bibliographystyle{JHEP}
\bibliography{bib}

\end{document}